\documentclass[apj,twocolumn,twocolappendix,numberedappendix]{openjournal}

\usepackage[dvipsnames]{xcolor}
\usepackage{hyperref}
\hypersetup{colorlinks=true,citecolor=blue,linkcolor=blue,urlcolor=blue,pdftitle={Identifying and Distinguishing Quenching Galaxies with Spatially Resolved Star Formation in Mock CASTOR and NGRST Observations},pdfauthor={Cameron Lawlor-Forsyth; Michael L. Balogh; Sean L. McGee; Gregory H. Rudnick},pdfsubject={}}
\usepackage{orcidlink}
\usepackage[hang]{footmisc}
\definecolor{doiPink}{RGB}{255, 0, 255}

\usepackage{enumitem}

\usepackage[largesc]{newtx}
\usepackage{couriers}
\DeclareSymbolFont{CMop}{OT1}{cmr}{m}{n}
\SetSymbolFont{CMop}{bold}{OT1}{cmr}{bx}{n}
\DeclareSymbolFont{CMlet}{OML}{cmm}{m}{it}
\SetSymbolFont{CMlet}{bold}{OML}{cmm}{b}{it}
\DeclareSymbolFont{CMSy}{OMS}{cmsy}{m}{n}
\SetSymbolFont{CMSy}{bold}{OMS}{cmsy}{b}{n}
\DeclareSymbolFont{AMSa}{U}{msa}{m}{n}
\SetSymbolFont{AMSa}{bold}{U}{msa}{m}{n}

\AtBeginDocument{
    \DeclareMathSymbol{/}{\mathord}{CMop}{"2F}%
    \DeclareMathSymbol{+}{\mathbin}{CMop}{"2B}%
    \DeclareMathSymbol{-}{\mathbin}{CMSy}{"00}%
    \DeclareMathSymbol{=}{\mathrel}{CMop}{"3D}%
    \DeclareMathSymbol{<}{\mathrel}{CMlet}{"3C}%
    \DeclareMathSymbol{>}{\mathrel}{CMlet}{"3E}%
    \DeclareMathSymbol{\leqslant}{\mathrel}{AMSa}{"36}
    \DeclareMathSymbol{\geqslant}{\mathrel}{AMSa}{"3E}%
    \DeclareMathSymbol{\lesssim}{\mathrel}{AMSa}{"2E}%
    \DeclareMathSymbol{\gtrsim}{\mathrel}{AMSa}{"26}%
    \DeclareMathSymbol{\sim}{\mathrel}{CMSy}{"18}%
    \DeclareMathSymbol{\pm}{\mathrel}{CMSy}{"06}%
    \DeclareMathSymbol{\approx}{\mathrel}{CMSy}{"19}%
    \DeclareMathSymbol{\times}{\mathbin}{CMSy}{"02}%
    \DeclareMathSymbol{\Delta}{\mathalpha}{CMop}{1}%
    \DeclareMathSymbol{\Phi}{\mathalpha}{CMop}{"08}%
    \DeclareMathSymbol{\Omega}{\mathalpha}{CMop}{10}%
    \DeclareMathSymbol{\alpha}{\mathalpha}{CMlet}{11}%
    \DeclareMathSymbol{\beta}{\mathalpha}{CMlet}{12}%
    \DeclareMathSymbol{\delta}{\mathalpha}{CMlet}{14}%
    \DeclareMathSymbol{\lambda}{\mathalpha}{CMlet}{"15}%
    \DeclareMathSymbol{\mu}{\mathalpha}{CMlet}{22}%
    \DeclareMathSymbol{\sigma}{\mathalpha}{CMlet}{27}%
}

\makeatletter
    \patchcmd{\NAT@citex} 
        {\@citea\NAT@hyper@{%
            \NAT@nmfmt{\NAT@nm}%
            \hyper@natlinkbreak{\NAT@aysep\NAT@spacechar}{\@citeb\@extra@b@citeb}%
            \NAT@date}}
        {\@citea
        \NAT@nmfmt{\NAT@nm}%
        \NAT@aysep\NAT@spacechar
        \NAT@hyper@{\NAT@date}}%
        {}{}
    \patchcmd{\NAT@citex} 
        {\@citea\NAT@hyper@{%
            \NAT@nmfmt{\NAT@nm}%
            \hyper@natlinkbreak{\NAT@spacechar\NAT@@open\if*#1*\else#1\NAT@spacechar\fi}%
            {\@citeb\@extra@b@citeb}%
            \NAT@date}}
        {\@citea
        \NAT@nmfmt{\NAT@nm}%
        \NAT@spacechar\NAT@@open\if*#1*\else#1\NAT@spacechar\fi
        \NAT@hyper@{\NAT@date}}%
        {}{}
\makeatother

\newcommand{\CC}{C\nolinebreak\hspace{-.05em}\raisebox{.4ex}{\tiny\bf +}\nolinebreak\hspace{-.10em}\raisebox{.4ex}{\tiny\bf +}}
\def\CC{{C\nolinebreak[4]\hspace{-.05em}\raisebox{.4ex}{\tiny\bf ++}}}

\newlength{\licenseiconwidth}
\newlength{\licenseicongap}
\newlength{\licenseindent}
\newcommand{\licensebox}{%
  \begin{figure}[!b]
    \footnotesize\linespread{1.2}\selectfont
    \setlength{\parindent}{0pt}
    \begingroup
      \parshape=3
        \licenseindent \dimexpr\columnwidth-\licenseindent\relax
        \licenseindent \dimexpr\columnwidth-\licenseindent\relax
        0pt \columnwidth
      \noindent
      \makebox[0pt][l]{%
        \hspace*{-\licenseindent}%
        \smash{%
          \raisebox{-1.1\baselineskip}[0pt][0pt]{%
            \includegraphics[width=\licenseiconwidth]{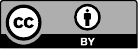}%
          }%
        }%
      }%
      Original content from this work may be used under the terms of the \href{https://creativecommons.org/licenses/by/4.0/}{Creative Commons Attribution 4.0 licence}. Any further distribution of this work must maintain attribution to the author(s) and the title of the work, journal citation and DOI.\par
    \endgroup
  \end{figure}
}

\journalinfo{The Open Journal of Astrophysics}
\historyline{Received 2026 July 17}
\shortauthors{Lawlor-Forsyth et al.}
\defcitealias{lawlor2026}{Paper~I}
\newcommand{\rotaci}{ROTAC et al.~\citeyearpar{rotac2025}}
\newcommand{\rotacp}{(ROTAC et al.~\citeyear{rotac2025})}
\newcommand{\rotact}{ROTAC et al.~\citeyear{rotac2025}}

\begin{document}

\title{Identifying and Distinguishing Quenching Galaxies with Spatially Resolved Star Formation in Mock CASTOR and NGRST Observations\vspace{-1.5cm}}

\author{Cameron~Lawlor-Forsyth$^{\text{\color{blue}1,2}}$\orcidlink{0000-0002-2958-0593}}
\author{Michael~L.~Balogh$^{\text{\color{blue}1,2}}$\orcidlink{0000-0003-4849-9536}}
\author{Sean~L.~McGee$^{\text{\color{blue}3}}$\orcidlink{0000-0003-3255-3139}}
\author{Gregory~H.~Rudnick$^{\text{\color{blue}4}}$\orcidlink{0000-0001-5851-1856}\vspace{0.1cm}}

\affiliation{$^{\text{1}}$Department of Physics and Astronomy, University of Waterloo, Waterloo, ON N2L 3G1, Canada; \href{mailto:clawlorforsyth@uwaterloo.ca}{clawlorforsyth@uwaterloo.ca}}
\affiliation{$^{\text{2}}$Waterloo Centre for Astrophysics, University of Waterloo, Waterloo, ON N2L 3G1, Canada}
\affiliation{$^{\text{3}}$School of Physics and Astronomy, University of Birmingham, Birmingham B15 2TT, UK}
\affiliation{$^{\text{4}}$Department of Physics and Astronomy, University of Kansas, Lawrence, KS 66045, USA}

\begin{abstract}
We present synthetic images of galaxies that are in the stages of star formation quenching for the Cosmological Advanced Survey Telescope for Optical and UV Research (CASTOR) and Nancy Grace Roman Space Telescope (NGRST), based on simulations coming from the IllustrisTNG suite, as processed using the stellar population synthesis library \textsc{galaxev}. We account for the effects of dust and various sources of noise to produce mock observations that should mirror real observations. Using these synthetic images, we fit photometric observations in binned circular annuli using \texttt{FAST++} and a flexible star formation history, and recover well the spatially resolved stellar mass and star formation rate. We thereby measure various indicators (morphological metrics) of spatially resolved star formation activity in the context of galaxy quenching. We find that we are able to distinguish quenching galaxies from a mass-matched control sample of normal star forming galaxies. We additionally find that we can distinguish various quenching mechanisms, where galaxies consistent with an inside-out quenching signature can be separated from galaxies that display an outside-in signature. Using machine learning techniques the accuracy of this classification is reliable, and the progress through the quenching episode can be estimated for the different populations of quenching galaxies. We make predictions for the abundance of the various quenching populations in proposed surveys for CASTOR and NGRST, and find that these surveys will enable the classifications of thousands of quenching galaxies out to intermediate redshifts, and more when considering higher redshifts.
\end{abstract}

\keywords{\href{http://astrothesaurus.org/uat/584}{Galaxy clusters (584)}; \href{http://astrothesaurus.org/uat/594}{Galaxy evolution (594)}; \href{http://astrothesaurus.org/uat/2040}{Galaxy quenching (2040)}; \href{http://astrothesaurus.org/uat/597}{Galaxy groups (597)}; \href{http://astrothesaurus.org/uat/621}{Galaxy stellar content (621)}; \href{http://astrothesaurus.org/uat/2016}{Quenched galaxies (2016)}}

\maketitle

\section{Introduction}
\setcounter{footnote}{4}

\licensebox

Star formation is the primary regulator of galaxy evolution in the Universe \citep[e.g.,][]{zhang2018}, and is well traced with recombination lines and infrared luminosity \citep[e.g.,][]{kennicutt1998,bell2003,burgarella2005,calzetti2007,calzetti2010,kennicutt2012,calzetti2013}. As well, ultraviolet radiation is one of the most straightforward tracers of star formation, and is mostly dominated by massive stars with short lifetimes \citep[e.g.,][]{vink2020}. Understanding the spatial and temporal evolution of star formation rate across cosmic environments is necessary to understand galaxy evolution and quenching \citep[e.g.,][]{madau2014}. Many mechanisms capable of quenching star formation in galaxies have been proposed including internal processes like active galactic nuclei feedback \citep[e.g.,][]{dimatteo2005,bower2006,croton2006,fabian2012}. This feedback can lead to a lack of young stars in the central regions of galaxies \citep[e.g.,][]{dubois2013}. In addition to internal effects, the environment in which a galaxy resides can also play a role in quenching \citep[e.g.,][]{peng2010,peng2012}. Ram pressure stripping \citep[e.g.,][]{gunn1972,abadi1999,balogh2000,poggianti2017,boselli2022}, starvation or strangulation \citep{larson1980,peng2015}, and gravitational harassment \citep{moore1996,moore1998} have all been proposed as environmental effects, and can act to suppress star formation in the outskirts of a galaxy. Given these potential internal and external influences, the morphological evolution of a galaxy undergoing a certain kind of quenching will present different spatial signatures \citep{zhang2019,lawlor2026}, as traced by the evolution of the young stellar population with ultraviolet observations.

Observations that spatially resolve the young stellar populations and can distinguish different quenching evolutionary signatures are clearly ideal, given the additional information included in such observations compared to integrated observations. In addition to large integral field unit surveys such as the Calar Alto Legacy Integral Field Area survey \citep[CALIFA;][]{sanchez2012}, the Sydney-AAO Multi-object Integral field spectrograph galaxy survey \citep[SAMI;][]{croom2012,bryant2015}, and Mapping Nearby Galaxies at Apache Point Observatory \citep[MaNGA;][]{bundy2015}, imaging studies using H$\alpha$ \citep[e.g.,][]{koopmann2004a,koopmann2004b,gavazzi2013,matharu2021,matharu2022} and spatially resolved spectral energy distribution fitting have also been completed \citep[e.g.,][]{abdurrouf2017,abdurrouf2018,jung2017,jafariyazani2019,morselli2019,nelson2021,abdurrouf2022,abdurrouf2023}. These studies have found galaxies with suppressed central star formation, indicative of a quenching mechanism that proceeds from the inside to the outskirts, in local \citep{abdurrouf2017,abdurrouf2022}, intermediate \citep{abdurrouf2018,jafariyazani2019}, and high redshift galaxies \citep{jung2017,abdurrouf2023}, and galaxies below the star forming main sequence \citep{morselli2019}. As well, \citet{nelson2021} found that massive galaxies below the main sequence \citep[e.g.,][]{brinchmann2004} also show centrally-suppressed star formation in both observations and simulations. Being able to distinguish different quenching signatures is therefore a powerful means of better understanding the shut down of star formation, with the previously mentioned studies all successfully employing spatially resolved spectral energy distribution fitting.

The next decade will be a transformational time in the field of astrophysics. New observatories such as the James Webb Space Telescope \citep[JWST;][]{gardner2006,gardner2023} and Euclid \citep{euclid2025} have already demonstrated the power of new facilities to probe previously inaccessible times in the history of the Universe \citep[e.g.,][]{castellano2022,finkelstein2022,finkelstein2023,naidu2022,adams2023,atek2023,donnan2023,harikane2023,labbe2023,labbe2025,kocevski2023,bezanson2024,greene2024,euclid2026a,euclid2026b,euclid2026c}. With the coming decade, new facilities such as the Vera C. Rubin Observatory \citep{ivezic2019}, in combination with its Legacy Survey of Space and Time \citep[LSST; e.g.,][]{lsst2009}, as well as the Nancy Grace Roman Space Telescope \citep[NGRST;][]{spergel2013,spergel2015,akeson2019} will similarly have a profound effect on our knowledge of the Universe, enabling large surveys of hundreds of millions of galaxies \citep[e.g.,][]{spergel2015,akeson2019,ivezic2019}. At the same time, aging facilities like the Hubble Space Telescope (HST) continue to be invaluable given the access to ultraviolet and blue-optical wavelengths \citep[e.g.,][]{cote2025,windhorst2025}, but will eventually be decommissioned. This will leave a notable gap in the coverage of the ultraviolet sky that is important for understanding young stellar populations within galaxies \citep[e.g.,][]{kennicutt1998,calzetti2013,madau2014}, before the launch of next generation facilities like the Habitable Worlds Observatory \citep{luvoir2019,gaudi2020}.

The Cosmological Advanced Survey Telescope for Optical and UV Research \citep[CASTOR;][]{cote2025} is a proposed mission to provide high-resolution imaging and spectroscopy over a ${\sim} 0.25~\text{deg}^{2}$ field of view at ultraviolet and blue-optical wavelengths ($0.15$--$0.55~\mu\text{m}$) simultaneously. With a planned launch for the end of the decade \citep{cote2023}, CASTOR will critically complement both space- and ground-based observatories that observe at longer wavelengths, such as Euclid, NGRST, and Rubin, and will prove instrumental in filling the aforementioned ``ultraviolet gap.'' The combination of these observatories will prove critical for understanding many astrophysical phenomena, including how galaxies have their star formation suppressed or quenched, given the ultraviolet coverage afforded by CASTOR.

In \citet[][hereafter \citetalias{lawlor2026}]{lawlor2026}, we showed that observationally motivated metrics that are based on the distribution of star formation can morphologically distinguish simulated quenching galaxies from star forming galaxies in IllustrisTNG \citep{nelson2018,pillepich2018b,springel2018}. We found that in many cases (but not all) the quenching population can be further divided into distinct quenching evolutionary signatures: a population where star formation is suppressed inside-out, and an outside-in population. For a minority of the quenching population this classification was not straightforward, as some galaxies showed the effects of both inside-out and outside-in quenching, and thus reside between the two previous populations. We considered this as a third class, that was ambiguously quenching. We found that the inside-out population in the simulation are more often primary$/$central galaxies found in the field, while outside-in quenched galaxies are commonly satellites in dense environments like galaxy groups and clusters, and begin quenching ${\sim} 1~\text{Gyr}$ after being accreted onto a larger halo. We found typical timescales associated with the duration of quenching, where the inside-out population can take up to ${\sim} 3.5~\text{Gyr}$ to quench at higher masses ($M_{*} \gtrsim 10^{11}~M_{\odot}$), but takes ${\sim} 2~\text{Gyr}$ to quench at lower masses ($M_{*} \lesssim 10^{11}~M_{\odot}$). The outside-in population has a typical quenching timescale of ${\sim} 1.5~\text{Gyr}$. In \citetalias{lawlor2026}, we additionally found that the morphological metrics for the quenching populations evolved differently through the quenching episode, and evolve distinctly compared to normal star forming galaxies that do not show evolution (by construction). We found that the morphological metrics can be used to estimate the progress through the quenching episode for the inside-out and outside-in populations.

However, in \citetalias{lawlor2026}, we did not make predictions or testable estimates for galaxies residing in the real Universe \citep[e.g.,][]{smith2018}. As such, in this paper we take a population of simulated quenched galaxies and create mock observations from which we can determine the feasibility of using the metrics introduced in \citetalias{lawlor2026} to find quenching galaxies in upcoming large galaxy surveys. As well, when creating our mock observations, we sample the quenching episode for each galaxy, thereby creating synthetic imaging that tracks the evolution of such systems, to understand the limits of this procedure in real data. Therefore, we artificially place our mock observations at a characteristic redshift that will be accessible to future galaxy surveys. Through this, we will be in a position to make predictions and testable estimates for the proportions of quenched galaxies with different morphological evolution in the real Universe, and provide constraints and suggestions for how to best identify such galaxies.

This paper is structured as follows: in Section~\ref{sec:methods}, we describe the simulated data that we have used previously in \citetalias{lawlor2026} along with the survey strategies for key upcoming large galaxy surveys, our procedure for creating mock observations based on those surveys, as well as the analysis pipeline for extracting physical measurements. In Section~\ref{sec:results}, we present our results and the recoverability of our metrics when using the mock observations as well as considerations for how to best adapt our metrics to real observations. In Section~\ref{sec:discussion}, we discuss our results and make predictions for upcoming surveys. Finally, in Section~\ref{sec:summary}, we present our summary and conclusion.

Throughout this paper we adopt a flat $\Lambda$CDM cosmology that is consistent with the TNG simulation \citep{weinberger2017,pillepich2018a}, based on the Planck intermediate results \citep{planck2016}: $H_{0} = 67.74~\text{km s}^{-1}~\text{Mpc}^{-1}$, $\Omega_{\text{m}} = 0.3089$, $\Omega_{\Lambda} = 0.6911$, and $\Omega_{\text{b}} = 0.0486$.

\section{Methodology}\label{sec:methods}

\subsection{IllustrisTNG and Galaxy Sample}\label{subsec:tng}

In \citetalias{lawlor2026}, we used the highest resolution run of the IllustrisTNG suite\footnote{\href{https://www.tng-project.org}{https:$//$www.tng-project.org}} \citep[e.g.,][]{nelson2018,nelson2019a,pillepich2018b,springel2018}, TNG50, to select simulated galaxies with $M_{*} \geqslant 10^{9.5}~M_{\odot}$ by $z = 0$. The IllustrisTNG simulations are cosmological gravo-magneto-hydrodynamical simulations and are the successor to the original Illustris \citep{genel2014,vogelsberger2014a,vogelsberger2014b} simulation. They include a revised galaxy formation model \citep{weinberger2017,pillepich2018a} compared to Illustris \citep{vogelsberger2013,torrey2014}. The moving mesh code \textsc{arepo} is used \citep[e.g.,][]{springel2010,weinberger2020}, alongside the \citet{chabrier2003} initial mass function \citep{pillepich2018a}, while a cosmology consistent with the \citet{planck2016} results is employed. The IllustrisTNG model aims to reproduce observations of the global star formation rate density (from $z = 8$ to the present), and galaxy sizes, halo gas fractions, the stellar mass--halo mass relation, the galaxy mass function, and the black hole--stellar mass relation \citep[all at $z = 0$;][]{rodriguez2019}. All simulation data from IllustrisTNG are publicly available \citep{nelson2019a}.

In \citetalias{lawlor2026}, the TNG50 simulation, that contains a cubic volume of $51.7~\text{Mpc}$ on a side \citep{nelson2019b,pillepich2019} and follows $2 \times 2160^{3}$ resolution elements (dark matter and baryons) and their evolution, was used as it provides high resolution for the baryonic elements \citep[average mass of $8.5 \times 10^{4}~M_{\odot}$;][]{nelson2019b,pillepich2019}. TNG50 has improved mass resolution by more than an order of magnitude compared to the flagship TNG100 simulation \citep{nelson2019b}. This improved mass resolution places TNG50 near modern ``zoom'' simulations but within a large cosmological volume \citep{nelson2019b}.

In \citetalias{lawlor2026}, we compared the star formation histories (SFHs) of quenched galaxies with those of normal star forming galaxies that reside on the star forming main sequence \citep[e.g.,][]{brinchmann2004,daddi2007,elbaz2007,noeske2007}, and that have a similar (${\leqslant} 0.1~\text{dex}$) stellar mass. We identified the quenched galaxies as those where their SFH was within the ${\pm} 2 \sigma$ limits of the main sequence population but then fell below the $2.5\text{th}$ percentile and stayed below that level. This procedure produced 361 quenched galaxies within TNG50, out of a possible 1666 galaxies with $M_{*} \geqslant 10^{9.5}~M_{\odot}$ by $z = 0$. We additionally identified the quenching episode for each of these galaxies by noting the last significant peak in star formation (relative to the entire SFH) as the onset, with the termination of quenching occurring when the $2.5\text{th}$ percentile is crossed. Please see \citetalias{lawlor2026} (e.g., their Figure~1) for full details on this procedure.

\begin{figure*}[t]
    \centering
    \includegraphics[width=\textwidth]{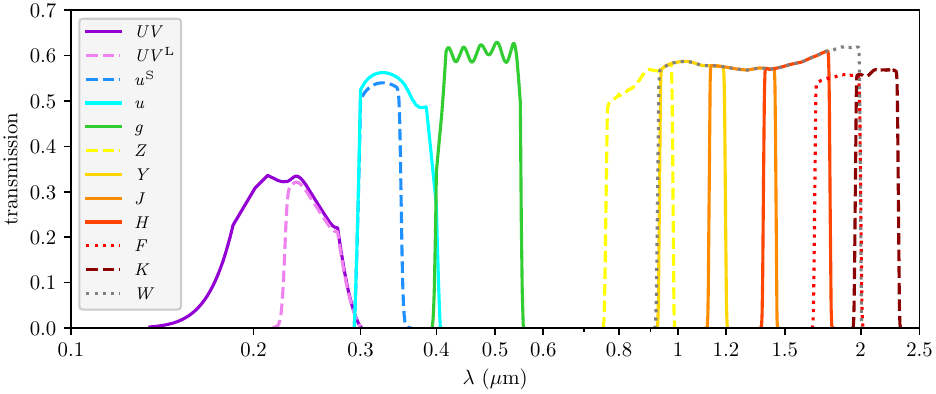}
    \caption{Filter transmission curves for the baseline CASTOR bands (\textit{UV}, \textit{u}, \textit{g}), the additional CASTOR bands after using the proposed broadband filter (\textit{UV}$^{\text{L}}$, \textit{u}$^{\text{S}}$), and the bands as part of the NGRST HLWAS Deep tier and Ultradeep component (\textit{Z}, \textit{Y}, \textit{J}, \textit{H}, \textit{F}, \textit{K}, and wide \textit{W}).}
    \label{fig:filters}
\end{figure*}

To summarize the sample used in \citetalias{lawlor2026}, by considering the 361 identified quenched galaxies through their respective quenching episodes, we reach a sample of 5365 quenching galaxies, with an additional 64,128 unique galaxies that represent all possible control star forming galaxies. These control star forming galaxies are similar to the stellar mass (${\leqslant} 0.1~\text{dex}$) of their respective quenching galaxy throughout the quenching galaxy's quenching episode, and so form a mass-matched sample. In total we have some 69,493 unique galaxies that we consider as our fiducial sample.

In order to test the feasibility of the metrics as used in \citetalias{lawlor2026}, we require mock observations against which we can compare our simulation-based results, that trace star formation and stellar mass. We therefore create mock observations with CASTOR and NGRST, as the ultraviolet and blue-optical observations from CASTOR will trace star formation, while the red-optical and near-infrared observations from NGRST will trace stellar mass.

\subsection{CASTOR Surveys}

Before describing the mock observation creation process, we first discuss the proposed CASTOR telescope and accompanying surveys that will provide spatially resolved images for tens of millions of galaxies \citep{cote2025}. In the following section we provide details for the proposed NGRST High-Latitude Wide-Area Survey (e.g., \citealt{dore2018}; \citealt{rotac2025}, hereafter \rotact).

CASTOR\footnote{\href{https://www.castormission.org}{https:$//$www.castormission.org}} is a proposed $1~\text{m}$ space telescope mission to provide high-resolution imaging and spectroscopy over a $0.247~\text{deg}^{2}$ field of view in three ultraviolet to blue-optical bandpasses, simultaneously \citep{cote2025}. These bandpasses cover the wavelength range from $0.15 - 0.55~\mu\text{m}$, where the image quality will have a point-spread function full width at half maximum of $0 \farcs 15$ \citep{cote2025}, comparable to the quality from HST \citep[and more than an order of magnitude better than the Galaxy Evolution Explorer;][]{martin2005}, with a native pixel size of $0 \farcs 1$. The design of CASTOR enables nearly diffraction-limited observations \citep{cote2025}, while additionally having excellent sensitivity ($5 \sigma$ point source detection at ${>} 27~\text{AB magnitude}$ in a $1000~\text{s}$ exposure) and a field of view roughly eighty times larger than HST's Advanced Camera for Surveys \citep{sirianni2005}. Compared to other proposed \citep[e.g., Ultraviolet Explorer;][]{kulkarni2021} and approved \citep[e.g., ULTRASAT;][]{sagiv2014} ultraviolet observatories, CASTOR will have superior image quality and sensitivity \citep{cote2025}, in addition to having a longer mission duration, making it an excellent discovery mission at ultraviolet and blue-optical wavelengths.

In addition to the main bandpasses\footnote{\href{https://github.com/castor-telescope/etc/tree/master/castor_etc/data/passbands}{https:$//$github.com$/$castor-telescope$/$etc$/$tree$/$master$/$castor\_etc$/$data$/$\\passbands}} (\textit{UV}, \textit{u}, \textit{g}) included in the CASTOR design, a proposed broadband filter is available that will split the \textit{UV}- and \textit{u}-channels\footnote{\href{https://github.com/castor-telescope/etc_notebooks/tree/master/data}{https:$//$github.com$/$castor-telescope$/$etc\_notebooks$/$tree$/$master$/$data}} \citep{cote2025}, enabling better spectral energy distribution sampling at short wavelengths. Filter transmission curves for the five available CASTOR bands are shown in Figure~\ref{fig:filters}.

Of the proposed legacy surveys that CASTOR will complete, there are three notable galaxy surveys of varying depth and area that comprise a tiered ``wedding cake'' approach \citep{cote2025,marshall2025}. The Wide Survey will image $2227~\text{deg}^{2}$ \citep{cote2025}, exactly overlapping with the NGRST High-Latitude Wide-Area Survey (described below), to a point source depth of ${\sim} 27.4~\text{mag}$ \citep{cote2025,marshall2025}. The Deep Survey will image $83~\text{deg}^{2}$ over six contiguous regions \citep{cote2023} that overlap with LSST$/$Euclid deep fields \citep{ivezic2019,euclid2022,euclid2025} and will reach a $5\sigma$ point source depth of $29.75~\text{mag}$ in the ultraviolet \citep{cote2025}. Finally, the Ultradeep Survey will image $1~\text{deg}^{2}$ over four pointings for ten times longer than the Deep Survey \citep{cote2023,cote2025}, reaching a depth ($5\sigma$, point source) of $31.12~\text{mag}$ \citep{cote2025} in the ultraviolet. For all CASTOR surveys, an initial observing sequence of \textit{UV}, \textit{u}, and {\textit{g}-band} images will be completed, followed by a subsequent sequence employing the broadband filter, producing \textit{UV}$^{\text{L}}$, \textit{u}$^{\text{S}}$, and {\textit{g}-band} images \citep{cote2023}. In Table~\ref{tab:surveys}, we present survey design parameters for the three CASTOR surveys.

\subsection{NGRST Surveys}

The Nancy Grace Roman Space Telescope\footnote{\href{https://science.nasa.gov/mission/roman-space-telescope}{https:$//$science.nasa.gov$/$mission$/$roman-space-telescope}} \citep[NGRST;][]{spergel2013,spergel2015,akeson2019} is an upcoming $2.4~\text{m}$ space telescope that will provide high-resolution imaging and spectroscopy over a $0.28~\text{deg}^{2}$ field of view across eight red-optical and near-infrared bands, through its Wide Field Instrument\footnote{\href{https://science.nasa.gov/mission/roman-space-telescope/wfi-technical}{https:$//$science.nasa.gov$/$mission$/$roman-space-telescope$/$wfi-technical}\label{footnote:wfi}} \rotacp. These bandpasses cover the wavelength range from $0.48 - 2.3~\mu\text{m}$, where the image quality will have a point-spread function full width at half maximum of $0 \farcs 058 - 0 \farcs 169$ (band-dependent), comparable to HST \rotacp, with a pixel size of $0 \farcs 11$. Only the two bluest bandpasses will not be diffraction-limited. NGRST will additionally have sensitivity comparable to HST, with $5 \sigma$ point source detection at $26.5~\text{AB magnitude}$ in a ${\sim} 600~\text{s}$ exposure \rotacp, and a field of view roughly ninety times larger than HST's Advanced Camera for Surveys \citep{sirianni2005}.

{\renewcommand{\arraystretch}{1.5}
\begin{deluxetable*}{cccccc}[t]
    \tablecaption{Survey design parameters for the CASTOR Wide, Deep, and Ultradeep Surveys, and the NGRST HLWAS tiers.\label{tab:surveys}}
    \tablehead{\colhead{Survey} & \colhead{Bands} & \colhead{Area} & \colhead{Exposure Time} & \colhead{$5\sigma$ Point Source Depth} & \colhead{Reference}\\[-0.1cm]
    \colhead{} & \colhead{} & \colhead{(deg$^{2}$)} & \colhead{(ks)} & \colhead{(AB mag)} & \colhead{}}
    \startdata
    CASTOR Wide & \textit{UV}, \textit{UV}$^{\text{L}}$, \textit{u}$^{\text{S}}$, \textit{u}, \textit{g} & 2227 & 1, 1, 1, 1, 2 & 27.57, 26.67, 27.03, 27.42, 27.41 & 1, 2, 3 \\
    CASTOR Deep & \textit{UV}, \textit{UV}$^{\text{L}}$, \textit{u}$^{\text{S}}$, \textit{u}, \textit{g} & 83 & 18, 18, 18, 18, 36 & 29.75, 28.90, 29.23, 29.37, 29.11 & 2, 3 \\
    CASTOR Ultradeep & \textit{UV}, \textit{UV}$^{\text{L}}$, \textit{u}$^{\text{S}}$, \textit{u}, \textit{g} & 1 & 180, 180, 180, 180, 360 & 31.12, 30.28, 30.61, 30.69, 30.38 & 2, 3 \\
    HLWAS Wide & \textit{H} & 2702 & 0.642 & 26.2 & 4 \\
    HLWAS Medium & \textit{Y}, \textit{J}, \textit{H} & 2415 & 0.642 & 26.5, 26.4, 26.4 & 4 \\
    HLWAS Deep & \textit{Z}, \textit{Y}, \textit{J}, \textit{H}, \textit{F}, \textit{K}, \textit{W} & 19.2 & 4.425 & 27.7, 27.7, 27.6, 27.5, 27.0, 25.9, 28.3 & 4 \\
    HLWAS Ultradeep & \textit{Y}, \textit{J}, \textit{H} & 5 & 8.85 & 28.2, 28.2, 28.1 & ~4 \vspace{0.1cm}
    \enddata
    \tablerefs{1. \citet{cote2023}, 2. \citet{cote2025}, 3. \citet{marshall2025}, 4. \rotaci}
\end{deluxetable*}}

NGRST will complete three core community surveys using WFI, one of which is the High-Latitude Wide-Area Survey\footnote{\href{https://science.nasa.gov/mission/roman-space-telescope/high-latitude-wide-area-survey-technical}{https:$//$science.nasa.gov$/$mission$/$roman-space-telescope$/$high-latitude-wide-area-survey-technical}}\textsuperscript{,}\footnote{\href{https://roman-docs.stsci.edu/roman-community-defined-surveys/high-latitude-wide-area-survey}{https:$//$roman-docs.stsci.edu$/$roman-community-defined-surveys$/$\\high-latitude-wide-area-survey}} (HLWAS; e.g., \citealt{dore2018}; \rotact). The HLWAS is composed of three tiers and an additional component with varying depth and area that similarly comprise a ``wedding cake'' approach, and includes both imaging and spectroscopy \rotacp. The Wide tier will cover $2702~\text{deg}^{2}$ in {\textit{H}-band}\footnote{We use \textit{Z}, \textit{Y}, \textit{J}, \textit{H}, \textit{F}, \textit{K}, and \textit{W} to refer to the NGRST F087, F106, F129, F158, F184, F213, and wide F146 filters, respectively, consistent with \rotaci.} imaging alone to a $5 \sigma$ point source depth of $26.2~\text{mag}$, while the Medium tier will image $2415~\text{deg}^{2}$ complementing the Wide tier in three-band \textit{Y}, \textit{J}, \textit{H} observations to a characteristic depth of $26.5~\text{mag}$ ($5 \sigma$, point source) in an exposure time of ${\sim}600~\text{s}$ \rotacp. The Deep tier will add \textit{Z}, \textit{F}, \textit{K}, and {\textit{W}-band} observations to the \textit{Y}, \textit{J}, \textit{H} imaging over $19.2~\text{deg}^{2}$ to reach $1.2~\text{mag}$ deeper across the seven bands, while $5~\text{deg}^{2}$ of the Deep tier will finally comprise the Ultradeep component, where \textit{Y}, \textit{J}, \textit{H} imaging is completed for an additional ${\sim} 0.5~\text{mag}$ deeper compared to the Deep tier \rotacp. We also note that according to \rotaci, the Deep tier and Ultradeep component are additive to the shallower tiers, enabling exposures of multiple hours for the \textit{Y}, \textit{J}, and {\textit{H}-bands}. The seven bands\footnote{We use the most recent transmission curves from 2024 March 27. For each filter, 18 sensor chip assemblies are available, each with their own respective bandpass quantum efficiencies. We determine a median bandpass across all sensor chip assemblies for each of the available bandpasses.} that will be used as part of the Deep tier and Ultradeep component have filter transmission curves shown in Figure~\ref{fig:filters}, while the survey design parameters for the three HLWAS survey tiers (and component) are presented in Table~\ref{tab:surveys}.

\subsection{Creating Mock Observations}\label{subsec:mocks}

To produce our mock observations, we predict the light distribution as determined using the simulation stellar particles and account for the effects of dust through the use of a spatially-varying foreground screen, given that IllustrisTNG does not account for dust \citep[e.g.,][]{rodriguez2019,bottrell2024}. While radiative transfer calculations would produce more accurate synthetic images, they are computationally expensive \citep[e.g.,][]{hayek2010}, and the extra precision afforded by radiative transfer is not warranted here. Indeed, we initially pursued using radiative transfer with the code SKIRT \citep{baes2011,camps2015}, and found that the results of interest (e.g., stellar mass, star formation rate) were not significantly affected. These findings are similar to those of \citet{rodriguez2019} who found that a simplified pipeline (described below) for producing mock observations creates images that are effectively indistinguishable from images produced with SKIRT for IllustrisTNG galaxies with low gas fractions.

\subsubsection{Stellar Populations}\label{subsubsec:galaxev}

To begin our analysis we take simulation cutouts (that include all stellar particles) for our sample galaxies and produce noise-free observations using the \textsc{galaxev} stellar population synthesis code \citep{bruzual2003}, following the prescription of \citet{rodriguez2019} and their \textsc{galaxev} pipeline\footnote{\href{https://archive.softwareheritage.org/browse/origin/https://github.com/vrodgom/galaxev_pipeline}{https:$//$github.com$/$vrodgom$/$galaxev\_pipeline}}: every galaxy is observed from a single viewing angle that is perpendicular to the simulation \textit{xy}-plane, and given the random orientations of the galaxies within the simulation volume, this produces random viewing angles across the sample. The field of view of each observation is ten times the (3D) stellar half-mass (effective) radius for a respective galaxy, ensuring the final images are large enough to facilitate a similar analysis as was completed in \citetalias{lawlor2026}. The number of pixels is chosen to match potential observations from CASTOR and NGRST after dithering (\citealt{cote2025}; \rotact), with a resulting pixel scale of $0 \farcs 05$. We output our synthetic images at a single characteristic intermediate redshift ($z = 0.5$), that is accessible to current and upcoming$/$proposed large galaxy surveys (e.g., Euclid, NGRST, LSST, and CASTOR). We discuss surface brightness and physical size effects below (see Section~\ref{sec:discussion}).

Following \citet{rodriguez2019}, we use the standard and preferred \citet{bruzual2003} simple stellar population models in \textsc{galaxev}, that use the ``{Padova~1994} library'' evolutionary tracks \citep{alongi1993,bressan1993,fagotto1994a,fagotto1994b,fagotto1994c,girardi1996} with a \citet{chabrier2003} initial mass function to model the spectra of the stellar particles coming from the simulation. These models describe the rest-frame luminosity per unit wavelength of a simple stellar population as a function of age, wavelength, and metallicity \citep{bruzual2003}. We use an updated version from 2016 of the \textsc{galaxev} code that provides better wavelength and metallicity coverage compared to its initial release\footnote{\href{https://www.bruzual.org/bc03/Updated_version_2016}{https:$//$www.bruzual.org$/$bc03$/$Updated\_version\_2016}}. These updated models are sampled at 221 unequally spaced ages from $0$ to $20~\text{Gyr}$, at 2023 unequally spaced wavelengths between $91~\Angstrom$ and $36~\text{mm}$, and at seven unequally spaced metallicities between $10^{-4}$ and $0.1$ \citep{bruzual2003}.

We then calculate the observer-frame flux per unit wavelength for each simple stellar population across all age and metallicity combinations, accounting for cosmological effects at our characteristic redshift (as described above). We use the filter response curves for the CASTOR \textit{UV}, \textit{UV}$^{\text{L}}$, \textit{u}$^{\text{S}}$, \textit{u}, \textit{g}, and NGRST \textit{Z}, \textit{Y}, \textit{J}, \textit{H}, \textit{F}, \textit{K}, and {\textit{W}-bands} to determine integrated apparent magnitudes as measured by an observer for each simple stellar population age$/$metallicity combination. This procedure creates a simple ``look-up'' table that can be interpolated and normalized using the age, metallicity, and initial birth mass of the stellar particle to produce apparent magnitudes for all stellar particles across all 12 CASTOR and NGRST bands \citep{rodriguez2019}. We then map and add the apparent magnitudes for all stellar particles for a given galaxy onto the final image grid, using the image size and resulting pixel scale as described above.

\subsubsection{Dust Modeling}\label{subsubsec:dust}

To account for the effects of dust, we employ a simple physically-motivated toy model that describes the spatial variation of a foreground dust screen. To motivate our parametrization for this toy model, we consider the work of \citet{greener2020}. In their work, \citet{greener2020} produced spatially resolved measurements of dust attenuation obtained from full-spectrum stellar population fits for nearly 140,000 spaxels in 232 star forming galaxies in MaNGA. Based on these fits, \citet{greener2020} determined the median dust attenuation as a function of radius, binned by galaxy stellar mass and offset from the star forming main sequence \citep[e.g.,][]{brinchmann2004}. They find that galaxies at all stellar masses show more attenuation in the central regions (${\lesssim} 0.5~R_{\text{e}}$), compared to outer regions (${\gtrsim} 1.5~R_{\text{e}}$), commonly reaching values of $A_{\text{V}} \approx 0.2~\text{mag}$ at large radii \citep{greener2020}. These median dust attenuation profiles appear consistent with a radially-declining exponential. In addition, higher-mass galaxies generally have more attenuation than lower-mass galaxies in their central regions, and galaxies closer to the star forming main sequence have attenuation profiles that are enhanced (i.e. a positive offset) compared to galaxies that are farther from the main sequence \citep{greener2020}.

With these results in mind, we construct our toy model: using the output from TNG50 in \citetalias{lawlor2026} (e.g., their Figure~2) we were able to determine the star forming main sequence as a function of time through all snapshots, where this relationship was described with a simple linear model following \citet{donnari2019} and \citet{pillepich2019}. Here, we use that model and determine the distance, $\Delta \text{MS}$, from the main sequence for every galaxy in our sample based on integrated stellar mass and star formation rate. We parameterize our spatially-varying dust law with a radially-declining exponential that is a function of $\Delta \text{MS}$ and the integrated stellar mass of the galaxy, $M_{*}$, as
\begin{equation}\label{eq:dust}
    A_{\text{V}} (M_{*}, \Delta \text{MS}) = Be^{-x} + C,
\end{equation}
where $A_{\text{V}}$ is the {\textit{V}-band} dust extinction, the coefficient ${B = \max(0.2, \log M_{*} - 9.5)}$, $x$ is the distance from the center of the image to the center of every pixel, and ${C = \max(0, 0.2 + 0.1 \Delta \text{MS})}$. As suggested above, the factors involved in $B$ and $C$ were determined empirically to produce dust extinction profiles that are qualitatively similar to \citet{greener2020}.

\begin{figure}[t]
    \centering
    \includegraphics[width=\columnwidth]{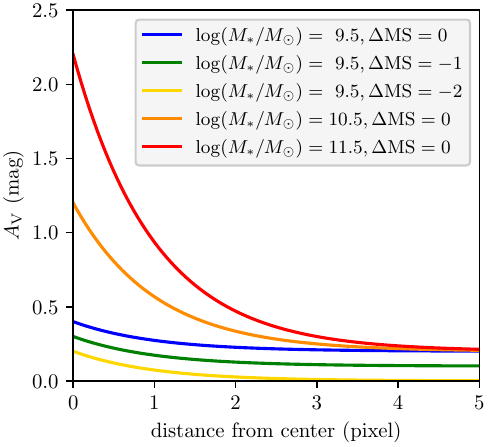}
    \caption{Example dust extinction profiles using our toy model from Equation~(\ref{eq:dust}), shown with different colors. At our chosen characteristic redshift of $z = 0.5$, one pixel corresponds with ${\sim} 0.31~\text{kpc}$.}
    \label{fig:dust}
\end{figure}

With the above parametrization, massive galaxies will have more dust in their central regions (from $B$), and galaxies closer to the main sequence will have more dust spread throughout the entirety of the stellar disk (from $C$). With this approach it is not necessarily critical that we reproduce the actual dust geometry for each galaxy, but rather that we can eventually recover the star formation rate distribution even with a nonuniform dust distribution. Example extinction profiles based on this process are shown in Figure~\ref{fig:dust}. With these profiles in hand, we use the dust extinction curves from \citet{calzetti2000} to determine the extinction as a function of wavelength, $A_{\lambda}$, for each pixel, and use the filter response curves to similarly determine the total extinction per pixel per band, before applying to the noiseless synthetic images.

\begin{figure*}
    \centering
    \includegraphics[height=0.95\textheight]{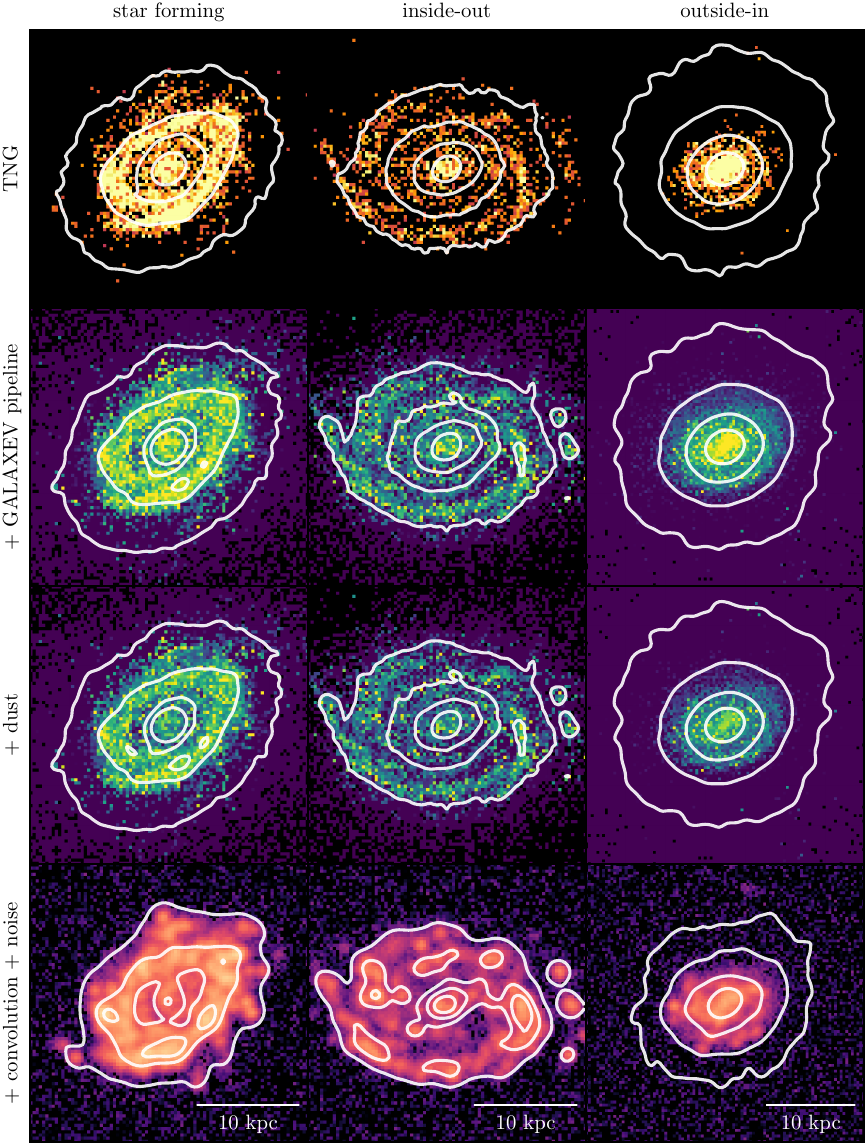}
    \caption{The mock observation image creation process for example star forming (left), inside-out (center), and outside-in (right) quenching galaxies, where the quenching galaxies are selected to be in the later stages of quenching. The top row displays a two dimensional projection of star formation as measured within the simulation ($100~\text{Myr}$ timescale), with contours shown in white that trace the distribution of stellar mass. The second row shows noiseless, unconvolved CASTOR \textit{UV} images after using the \textsc{galaxev} pipeline (see Section~\ref{subsubsec:galaxev}), along with contours from the NGRST \textit{H}-band image. The third row shows the effect of our foreground dust treatment (Section~\ref{subsubsec:dust}). The fourth row shows the final mock observations after convolution and accounting for noise (Sections~\ref{subsubsec:optics} and \ref{subsubsec:noise}). In the bottom row, we show scale bars that describe the spatial scale of the images at $z = 0.5$.}
    \label{fig:mock_images}
\end{figure*}

We show example noiseless unconvolved images for three galaxies in Figure~\ref{fig:mock_images}, using the mock CASTOR and NGRST photometry. In the top row we show the 2D projection of star formation (on a $100~\text{Myr}$ timescale) as seen in TNG along with contours that describe the stellar mass. In the second row we show the example noiseless, unconvolved images using the mock observation process described above without dust, for the same galaxies, where the CASTOR {\textit{UV}-band} image is shown with contours from the NGRST {\textit{H}-band}. In the third row we show the same images with the effects of dust, that is strongest and most noticeable in the central regions of each galaxy. Moving from left to right, we show an example normal star forming galaxy, an inside-out quenching galaxy, and an outside-in quenching galaxy, both in the later stages of quenching (${\sim}$75\% of the way through the quenching episode).

\subsubsection{Telescope Optics}\label{subsubsec:optics}

Before we arrive at final mock observations, we must account for the effects of both a point spread function and observing noise.

Astronomical images are affected by the optical system of the telescope carrying out the observations. These effects can be adequately described through convolution with a point spread function (PSF) that matches the resolution of the proposed surveys. For CASTOR, the expected image quality will have a PSF full width at half maximum (FWHM) of $0 \farcs 15$ for the \textit{g}-band \citep{cheng2024,cote2025}, and the NGRST HLWAS will have resolutions of $0 \farcs 058 - 0 \farcs 169$ (band-dependent; \rotact), where only the {\textit{K}-band} will have a PSF FWHM larger than $0 \farcs 15$. In order to produce PSF-matched mock observations, for simplicity we choose an azimuthally-symmetric Gaussian with a FWHM of $0 \farcs 15$ for all of our synthetic images, thereby matching the expected CASTOR image quality. In the fourth row of Figure~\ref{fig:mock_images}, we show our images after convolution with the PSF.

\subsubsection{Noise Modeling}\label{subsubsec:noise}

We next add appropriate noise (background, detector, and read) to our synthetic images. Following \citet{cheng2024}, background values for CASTOR were based on the HST exposure time calculator\footnote{\href{https://etc.stsci.edu}{https:$//$etc.stsci.edu}}, and include contributions from earthshine, zodiacal light, and geocoronal emission lines in the ultraviolet. We adopt the background values for CASTOR from the CASTOR exposure time calculator\footnote{\href{https://github.com/CASTOR-telescope/ETC}{https:$//$github.com$/$CASTOR-telescope$/$ETC}} \citep{cheng2024} for average earthshine and low zodiacal background, as the CASTOR exposure time calculator does not account for positional variations in zodiacal light. Background values for the CASTOR filters are listed in Table~\ref{tab:filters}.

Background values for NGRST were based on JWST backgrounds, and include contributions from zodiacal light and the galactic interstellar medium, where the zodiacal light model was based on the model of \citet{wright1998} and \citet{gorjian2000}, and is consistent with the model of \citet{reach1997} and \citet{kelsall1998} at the wavelengths under consideration. Emission from the diffuse galactic interstellar medium was based on the models of \citet{schlegel1998} and \citet{arendt1998}. These two components have been in use since the Spitzer Space Telescope \citep{werner2004} and Herschel Space Observatory \citep{pilbratt2010}, and continue with Euclid\footnote{\href{https://irsa.ipac.caltech.edu/applications/BackgroundModel}{https:$//$irsa.ipac.caltech.edu$/$applications$/$BackgroundModel}}. The output backgrounds are also consistent with that from the JWST Backgrounds Tool\footnote{\href{https://jwst-docs.stsci.edu/jwst-other-tools/jwst-backgrounds-tool}{https:$//$jwst-docs.stsci.edu$/$jwst-other-tools$/$jwst-backgrounds-tool}} v1.3.0 \citep[e.g.,][]{rigby2023}. We produce background estimates for two Rubin Deep Drilling Fields\footnote{\href{https://survey-strategy.lsst.io/baseline/ddf.html}{https:$//$survey-strategy.lsst.io$/$baseline$/$ddf.html}} \citep[e.g.,][]{ivezic2019} that are the locations of the NGRST HLWAS Deep tier \rotacp: COSMOS \citep{scoville2007} and XMM-LSS \citep{pierre2004}. These fields are additionally proposed locations for the CASTOR Deep and Ultradeep surveys \citep{cote2023}. For each field we take the median zodiacal value over the times of the year where the field is in a characteristic spacecraft viewing zone. We then average the output backgrounds for the two fields, producing a final background estimate as a function of wavelength. Background values for the NGRST filters are listed in Table~\ref{tab:filters}.

{\renewcommand{\arraystretch}{1.5}
\begin{deluxetable}{cccccc}[t]
    \tablecaption{Noise parameters for the CASTOR and NGRST filters.\label{tab:filters}}
    \tablehead{\colhead{Filter} & \colhead{$\lambda$} & \colhead{Background} & \colhead{Read} & \colhead{Dark} & \colhead{Ref.}\\[-0.1cm]
    \colhead{} & \colhead{($\mu\text{m}$)} & \colhead{($\text{mag}~\text{asec}^{-2}$)} & \colhead{($\text{e}^{-}~\text{px}^{-1}$)} & \colhead{($\text{e}^{-}~\text{s}^{-1}~\text{px}^{-1}$)} & \colhead{}}
    \startdata
    \textit{UV} & 0.226 & 26.09 & 3.0 & 0.0001 & 1, 2 \\
    \textit{UV}$^{\text{L}}$ & 0.252 & 25.32 & 3.0 & 0.0001 & 1, 2 \\
    \textit{u}$^{\text{S}}$ & 0.323 & 24.96 & 3.0 & 0.0001 & 1, 2 \\
    \textit{u} & 0.345 & 24.53 & 3.0 & 0.0001 & 1, 2 \\
    \textit{g} & 0.475 & 22.96 & 3.0 & 0.0001 & 1, 2 \\
    \textit{Z} & 0.872 & 21.88 & 13.81 & 0.018 & 3 \\
    \textit{Y} & 1.062 & 21.98 & 13.81 & 0.018 & 3 \\
    \textit{J} & 1.284 & 22.15 & 13.81 & 0.018 & 3 \\
    \textit{H} & 1.580 & 22.32 & 13.81 & 0.018 & 3 \\
    \textit{F} & 1.838 & 22.47 & 13.81 & 0.018 & 3 \\
    \textit{K} & 2.131 & 22.64 & 13.81 & 0.018 & 3 \\
    \textit{W} & 1.434 & 22.19 & 13.81 & 0.018 & ~3 \vspace{0.1cm}
    \enddata
    \tablerefs{1. \citet{cheng2024}, 2. \citet{cote2025}, 3. See footnote~\ref{footnote:wfi}}
\end{deluxetable}}

For both CASTOR and NGRST, we adopt projections for read noise and dark current from \citet{cheng2024} and \citet{cote2025} for CASTOR, and from the WFI Technical page for NGRST\footnote{See footnote~\ref{footnote:wfi}.}. These values are listed in Table~\ref{tab:filters}. We similarly adopt the respective survey strategies for CASTOR \citep{cote2023,cote2025,marshall2025} and the HLWAS \rotacp, to determine the number of CCD reads for each survey. We use exposure times equivalent to the Ultradeep survey for CASTOR, and the Ultradeep component within the Deep tier for the NGRST HLWAS (see Table~\ref{tab:surveys}). The exposure times and depths listed in Table~\ref{tab:surveys} for the CASTOR Ultradeep survey are the formal limits. A careful assessment of the in-orbit performance of the telescope will determine systematic factors, but we nonetheless use the exposure times as listed in Table~\ref{tab:surveys} for simplicity. For each CASTOR filter this is an exposure time of $50~\text{hr}$, except the {\textit{g}-band} that has a total exposure of $100~\text{hr}$ \citep{cote2023,cote2025,marshall2025}. The improved depth of the Ultradeep survey enables us to probe galaxies with levels of star formation that could prove difficult to detect with the Wide survey alone. For the HLWAS filters in the Deep tier this is $4.425~\text{ks}$ for the \textit{Z}, \textit{F}, \textit{K}, and wide {\textit{W}-bands}, while the \textit{Y}, \textit{J}, and {\textit{H}-bands} in the Ultradeep component have total exposure times of $13.917~\text{ks}$ due to the extra (additive) Medium and Deep tiers \rotacp.

For each filter, we add the contributions from the background and dark current to our PSF-convolved noiseless synthetic images before sampling from a Poisson distribution. We next add the RMS noise values by sampling from a Gaussian where the standard deviation is determined by the number of reads and the read noise. We finally subtract the background to produce mock observations that include the effects of the telescope optical system and noise. We also produce images that describe the uncertainty in these images, from which we can easily compute signal-to-noise ratios. In the bottom row of Figure~\ref{fig:mock_images}, we show our final mock observations after accounting for noise, assuming the exposure times as discussed above.

\subsection{Spectral Energy Distribution Fitting}\label{subsec:sed}

\begin{figure*}
    \centering
    \includegraphics[width=\textwidth]{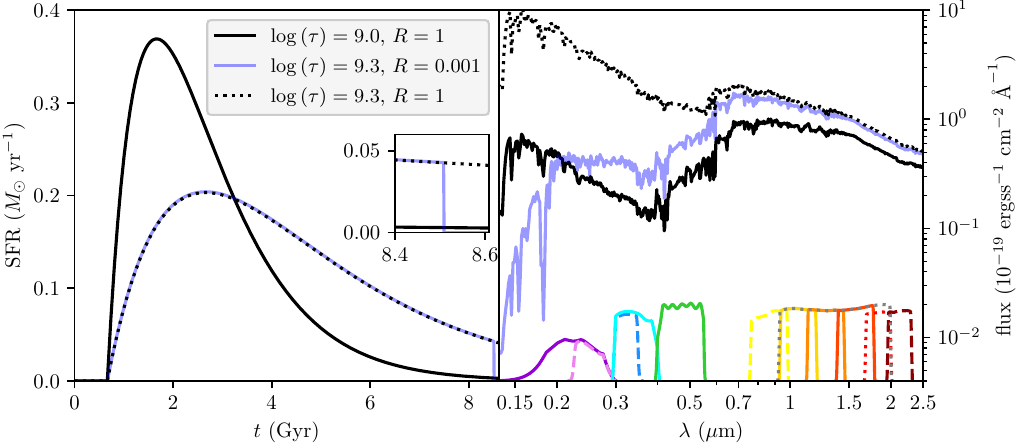}
    \caption{Left: example star formation histories for galaxies at $z = 0.5$ showing different combinations of the $e$-folding time $\tau$ and the suppression$/$burst coefficient $R$, using the star formation history as described in Equation~(\ref{eq:sfh}). The inset shows a zoom into the last ${\sim} 200~\text{Myr}$, where the final $100~\text{Myr}$ suppression for the star formation history with $\log{(\tau/\text{Gyr})} = 9.3, R = 10^{-3}$ is evident. Right: the resulting spectral energy distributions, color-matched to the star formation histories on the left, shown without the dimming effects of a foreground dust screen. Also shown in the lower part of the panel are the filter transmission curves for CASTOR and NGRST, color-coded the same as in Figure~\ref{fig:filters}.}
    \label{fig:sfhs_seds}
\end{figure*}

With our final mock observations produced for our sample of star forming and quenching$/$quenched galaxies, we produce photometric tables for use with stellar population synthesis fitting software to fit spectral energy distributions (SEDs). To best reproduce the radial profiles used in \citetalias{lawlor2026}, we measure the photometry in 20 equally-spaced (in radius) concentric circular annuli out to five times the effective radius (or half mass radius, as measured in TNG) for each galaxy. We then fit this photometry using the SED fitting code \texttt{FAST++}\footnote{\href{https://github.com/cschreib/fastpp}{https:$//$github.com$/$cschreib$/$fastpp}} \citep[e.g.,][]{schreiber2018a}, which is an updated {\CC} version of the original code presented in \citet{kriek2009}.

In our fitting we use the same \textsc{galaxev} stellar population synthesis model \citep{bruzual2003} as was used to produce the mock observations with the updates described above, namely, the standard {Padova~1994} evolutionary tracks with the better wavelength ($91~\Angstrom$ to $36~\text{mm}$) and metallicity ($0.008$ to $0.05$) coverage. We use a \citet{chabrier2003} initial mass function and a \citet{calzetti2000} dust attenuation law, once again matching the process used when producing the mock observations. We likewise use the characteristic redshift described above to produce the mock observations as a spectroscopic redshift. For our star formation history (SFH), we use a flexible hybrid SFH that is based on a delayed-tau star formation history \citep[e.g.,][]{moustakas2013,speagle2014,ciesla2017,carnall2019,johnson2021} until $100~\text{Myr}$ ago (in lookback time). In the final $100~\text{Myr}$ we assume the same delayed-tau model, but with a multiplicative coefficient, $R$, that controls the overall scaling \citep[e.g.,][]{fumagalli2011,ciesla2016,ciesla2018,ciesla2021,schreiber2018b}. Mathematically, this SFH is described as
\begin{equation}\label{eq:sfh}
    \text{SFR}(t) \propto \begin{cases}
        R \times (t - t_{\text{f}}) e^{-(t - t_{\text{f}})/\tau}, & t_{\text{lb}} \leqslant \Delta t_{\text{SF}}, \\
        (t - t_{\text{f}}) e^{-(t - t_{\text{f}})/\tau}, & t_{\text{lb}} > \Delta t_{\text{SF}} \cap t \geqslant t_{\text{f}}, \\
        0, & t < t_{\text{f}},
    \end{cases}
\end{equation}
where $t$ is time, $t_{\text{f}}$ is the time when the first stars were formed, $\tau$ is the $e$-folding time of the stellar population, $t_{\text{lb}}$ is the lookback time, and $\Delta t_{\text{SF}} = 100~\text{Myr}$. The respective times are related by $t + t_{\text{lb}} = t_{\text{age}}$, where $t_{\text{age}}$ is the age of the stellar population, and is related to the formation time by $t_{\text{f}} + t_{\text{age}} = t_{0}$, where $t_{0}$ is the current age of the Universe.

\begin{figure*}
    \centering
    \includegraphics[width=\textwidth]{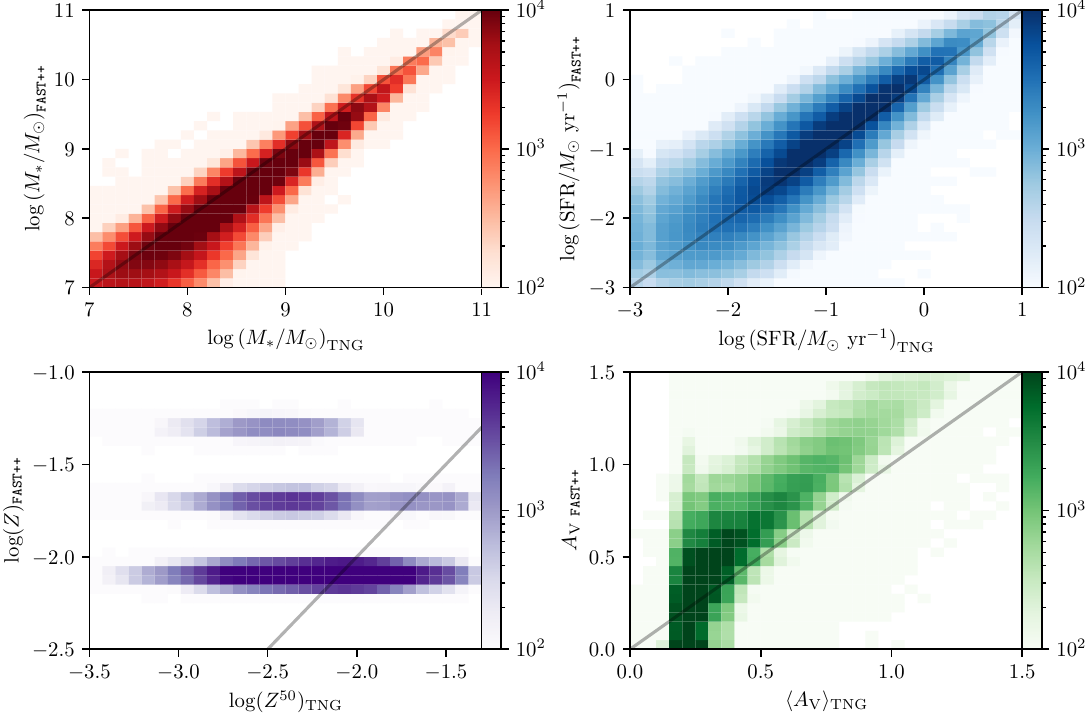}
    \caption{Comparisons of the fitted results from \texttt{FAST++} as a function of the simulated properties from TNG. Where relevant we show the logarithm for the given quantities, as noted in the axes labels. In all panels, we show a line of equality (gray). For the fitted metallicity and dust from \texttt{FAST++}, we perturb the values slightly for visualization purposes.}
    \label{fig:comparison}
\end{figure*}

We find that this SFH can capture both the broad scale evolution of long-lived stellar populations over long timescales, and also provide the flexibility necessary to describe rapid bursts or truncations in star formation over the most recent $100~\text{Myr}$ epoch, similar to \citet{ciesla2016,ciesla2018}. We list the parameter space investigated by the components of the SED fitting procedure in Table~\ref{tab:fitting}, while in Figure~\ref{fig:sfhs_seds}, we show example SFHs and resulting SEDs for different combinations of $\tau$ and $R$, where the ultraviolet to near infrared coverage afforded by CASTOR and NGRST enables us to distinguish SEDs with high confidence.

{\renewcommand{\arraystretch}{1.5}
\begin{deluxetable}{ccccc}
    \tablecaption{Spectral energy distribution fitting model parameters.\label{tab:fitting}}
    \tablehead{\colhead{Parameter} & \colhead{Minimum} & \colhead{Maximum} & \colhead{Step} & \colhead{$N_{\text{steps}}$}}
    \startdata
    $\log{(t_{\text{age}}/\text{Gyr})}$ & 8.8 & 9.9 & 0.1 & 12 \\
    $\log{(\tau/\text{Gyr})}$ & 8 & 10.5 & 0.1 & 26 \\
    $\log{(R)}$ & $-$3 & 2 & 0.1 & 51 \\
    $A_{\text{V}}$ (mag) & 0 & 1.8 & 0.1 & 19 \\
    $Z$ & 0.008 & 0.05 & irr. & 3 \\
    $z$ & 0.5 & 0.5 & & 1 \\
    \hline
    stellar populations & \multicolumn{4}{c}{\citet{bruzual2003}} \\
    initial mass function & \multicolumn{4}{c}{\citet{chabrier2003}} \\
    dust attenuation & \multicolumn{4}{c}{\citet{calzetti2000}} \vspace{0.1cm}
    \enddata
\end{deluxetable}}

In Figure~\ref{fig:comparison}, we show two dimensional histograms that compare our SED-fitted parameters with similar values either available directly in TNG, or derivable from TNG. The key parameters (per annulus) we consider are stellar mass, star formation rate, metallicity, and dust. For the metallicity per annulus in TNG, we take the median metallicity across all relevant stellar particles, while for the expected dust used for TNG, we use the average dust value across all pixels that comprise an annulus, using the foreground dust methodology described above. In the following, we take the difference of the fitted values from the expected (i.e. TNG) values, and describe the ``recovery'' of fitted values using the median difference and the standard deviation (scatter, $\sigma$) of the differences.

We find very good recovery when considering stellar mass, where the median difference and scatter are $0.19~\text{dex}$ and $0.23~\text{dex}$, respectively, when considering all masses, and $0.18~\text{dex}$ and $0.14~\text{dex}$ for $M_{*} \geqslant 10^{9}~M_{\odot}$. We next find very good recovery for star formation rate, with generally low scatter at higher values of star formation, and increasing scatter as star formation decreases, as expected. The median difference and scatter are $-0.13~\text{dex}$ and $0.46~\text{dex}$ across all star formation rates, and $-0.11~\text{dex}$ and $0.31~\text{dex}$ above $\text{SFR} \geqslant 0.1~M_{\odot}~\text{yr}^{-1}$. These results are encouraging as these parameters are critical for our next steps (see below). For the metallicity, we find poorer recovery of the expected metallicity: median difference and scatter of $-0.10~\text{dex}$ and $0.37~\text{dex}$, respectively, for all metallicities, and $0.21~\text{dex}$ and $0.15~\text{dex}$ above $Z = 0.01$. In general, \texttt{FAST++} finds best fit solutions with higher levels of metallicity compared to the median metallicity in TNG. This is partly due to the limited metallicities available in \textsc{galaxev} that we use within \texttt{FAST++}. We find that small changes in metallicity (i.e. changing from $Z = 0.008$ to $Z = 0.02$, or from $Z = 0.02$ to $Z = 0.05$, etc.) do not produce significant changes in the recovered SED, and additionally, the differences in metallicity are typically compensated with differences in dust. As a consequence, small changes in metallicity produce small variations in the recovered star formation rate at the level of ${\leqslant} 0.3~\text{dex}$ for ${>}$85\% of cases. Finally, for the strength of the foreground dust screen, we additionally find good recovery of the average dust screen (as described above), where the median difference and scatter are $-0.08~\text{mag}$ and $0.19~\text{mag}$ across all dust values, with a feature at $\langle A_{\text{V}} \rangle_{\text{TNG}} \approx 0.2~\text{mag}$ where \texttt{FAST++} can find a best fit solution that has additional dust. This feature is not seen anywhere else, and is a bias that is apparent at low signal-to-noise ratio (SNR). The nature of our constructed dust profiles mean they often reach ${\sim} 0.2~\text{mag}$ in the outer regions of a galaxy where signal-to-noise ratio is low. When restricting to annuli with \textit{H}-band $\text{SNR} \geqslant 3$, we find that this feature is significantly diminished and there is reduced scatter as well: $0.16~\text{mag}$, along with a median difference of $-0.09~\text{mag}$. Further increasing the lower limit on allowed \textit{H}-band SNR results in decreasing scatter.

As a sanity check, we additionally investigate the integrated values for stellar mass and star formation rate. When comparing the values from the simulation to the fitted results, we find median offsets of $0.18~\text{dex}$ and $-0.12~\text{dex}$, along with ${\pm} 1 \sigma$ ranges of $0.05~\text{dex}$ and $0.07~\text{dex}$ for stellar mass and star formation rate, respectively. These integrated values also function as a consistency check for the fits from the annuli, given that the annuli are fitted independently from the other annuli that comprise a galaxy. If the individual annular fits showed systematically different results we would expect to find larger differences and scatter between the integrated stellar masses and star formation rates.

\begin{figure*}
    \centering
    \includegraphics[width=\textwidth]{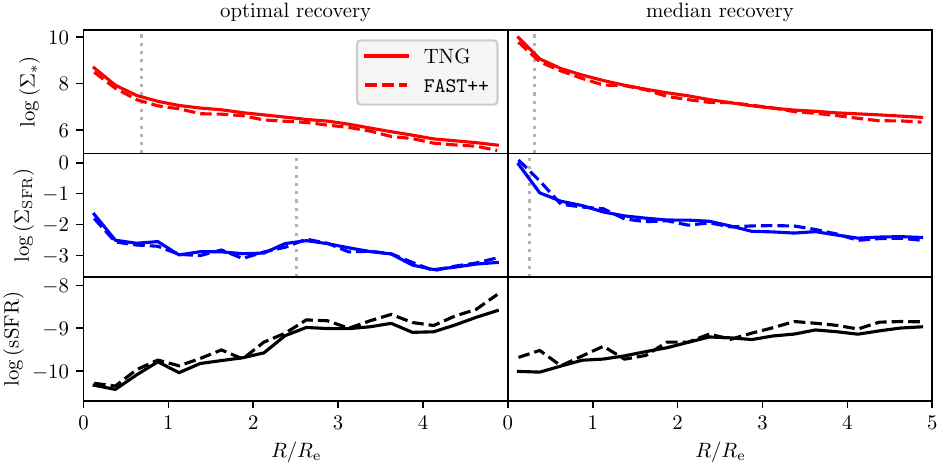}
    \caption{Example radial profiles showing recovered properties (dashed lines) when compared to the simulation (solid lines). Moving from top to bottom, we show the stellar mass radial profile (red, units of $M_{\odot}~\text{kpc}^{-2}$), the star formation rate radial profile (blue, units of $M_{\odot}~\text{yr}^{-1}~\text{kpc}^{-2}$), and the specific star formation rate radial profile (black, units of $\text{yr}^{-1}$). The galaxy on the left illustrates an ideal case where the respective radial profiles have optimal recovery when compared to their simulation counterparts, while the galaxy on the right shows a typical case, where the recovered profiles demonstrate what the median recovery is like. For the stellar mass and star formation rate profiles, we show the radius that encloses half of the stellar mass, or young stellar mass (see Section~\ref{subsubsec:adaptations}), respectively, as dotted vertical gray lines.}
    \label{fig:profiles}
\end{figure*}

We use the SED fitting results to produce radial profiles for star formation rate (SFR) and stellar mass, from which we can compute specific star formation rate ($\text{sSFR} = \text{SFR}/M_{*}$) radial profiles. For all galaxies we compute $\chi^{2}$ by comparing the fitted profiles to the profiles from the simulation for both the (log) stellar mass and star formation rate. These $\chi^{2}$ distributions are log-normal, with a median $\chi^{2}$ of $0.13^{+0.30}_{-0.08}$ and $3.1^{+14.0}_{-2.3}$ for stellar mass and star formation rate, respectively. By considering both values simultaneously, we determine an ideal case with optimal recovery (i.e. $\chi^{2} \approx 0$) and a typical case where the recovery is close to the median $\chi^{2}$ value. We show example radial profiles in Figure~\ref{fig:profiles}, for an ideal case (left), and a typical case (right).

\subsection{Morphological Metrics}

In \citetalias{lawlor2026}, we showed that the concentration of star formation, the size of the star forming disk relative to the stellar mass disk, and characteristic radii that trace sharp truncations of star formation can effectively discriminate various quenching mechanisms for galaxies undergoing a shut down of star formation, while also distinguishing these galaxies from a control sample of normal star forming galaxies. We refer the reader to \citetalias{lawlor2026} for full details. We include the formulations and brief descriptions for these parameters as used on the simulated galaxies below for completeness:
\begin{align}
    C_{\text{SF}}^{\text{TNG}} &= \text{SFR}_{< 1~\text{kpc}}/\text{SFR}_{\text{total}}, \\
    R_{\text{SF}}^{\text{TNG}} &= \log{(R_{\text{e, SF}}/R_{\text{e}})}, \\
    R_{\text{inner}}^{\text{TNG}} &= \log[\text{sSFR}(r)] \leqslant -10.5 \cap \text{d} (\text{sSFR})/\text{d}r \geqslant 1,\label{eq:Rinner} \\
    R_{\text{outer}}^{\text{TNG}} &= \log[\text{sSFR}(r)] \leqslant -10.5 \cap \text{d} (\text{sSFR})/\text{d}r \leqslant -1.\label{eq:Router}
\end{align}
The concentration of star formation, $C_{\text{SF}}^{\text{TNG}}$, is measured by comparing the star formation rate within the central $1~\text{kpc}$ relative to the total star formation rate integrated over the entire galaxy, while the size of the star forming disk relative to the stellar mass disk, $R_{\text{SF}}^{\text{TNG}}$, is determined by comparing the half mass radius (or effective radius) of young (${<} 100~\text{Myr}$) stellar particles with the effective radius of all stellar particles. The subsequent morphological metrics, the inner and outer truncation radii, $R_{\text{inner}}^{\text{TNG}}$ and $R_{\text{outer}}^{\text{TNG}}$, describe sharp truncations or abrupt changes in slope in the specific star formation rate radial profile, and are determined by moving outward from $0~R_{\text{e}}$ to $5~R_{\text{e}}$ for both. The metrics are then the outermost and innermost such radii that satisfy Equation~(\ref{eq:Rinner}) and (\ref{eq:Router}), respectively.

\begin{figure*}[t]
    \centering
    \includegraphics[width=\textwidth]{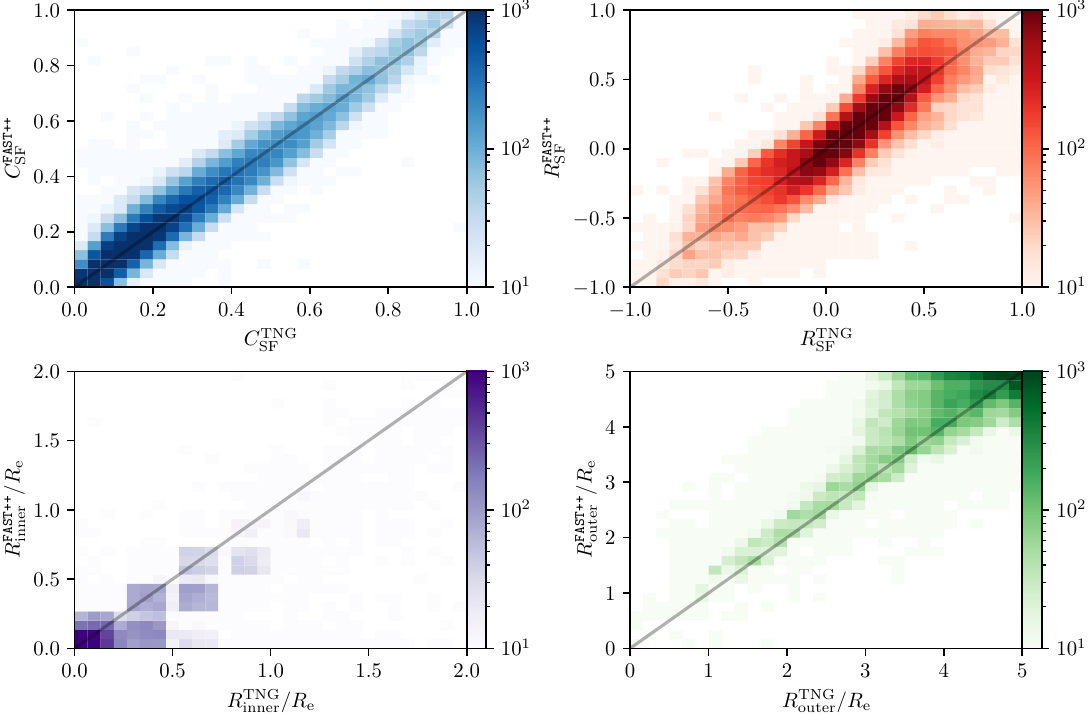}
    \caption{Comparisons of the observational proxies of the morphological metrics using the fitted results from \texttt{FAST++} as a function of the morphological metrics as computed using TNG. In all panels, we show 2D histograms where the color scale describes the density of points, and we additionally show a line of equality (gray). For the inner truncation radius, we zoom the panel into the relevant region for clarity purposes.}
    \label{fig:metric_comparison}
\end{figure*}

As described previously, with the above metrics, in \citetalias{lawlor2026} we showed that quenching galaxies in TNG50 have spatial signatures that are consistent with an inside-out quenching signature, an outside-in quenching signature, and an ambiguous signature, for some 361 quenched galaxies out of a possible 1666 galaxies with $M_{*} \geqslant 10^{9.5}~M_{\odot}$.

\subsubsection{Metric Adaptations}\label{subsubsec:adaptations}

To best recover observational proxies for these same metrics, we find that we must make adaptations for the metrics. For the concentration of star formation, $C_{\text{SF}}^{\texttt{FAST++}}$, we use the star formation rate radial profile and multiply with the characteristic timescale of $100~\text{Myr}$ to produce a young stellar mass profile. With this young mass profile, we then produce a cumulative young mass profile and interpolate to the radius corresponding to $1~\text{kpc}$ (projected), and compare with the final point (i.e. the outermost; $R \approx 5~R_{\text{e}}$) in the cumulative profile, taking the ratio as the observational proxy $C_{\text{SF}}^{\texttt{FAST++}}$. We describe alternative adaptations that we previously explored in Appendix~\ref{app:alternatives}.

For the size of the star forming disk relative to the stellar mass disk, $R_{\text{SF}}^{\texttt{FAST++}}$, we similarly produce a cumulative young mass profile following the steps described above, and additionally produce a cumulative stellar mass profile. For both cumulative profiles we interpolate to the point containing fifty percent of the profile, producing proxies for the effective radii for the star formation and stellar mass. For visualization purposes we show examples of these radii as vertical dotted gray lines for the two example galaxies in Figure~\ref{fig:profiles}. Finally we take the logarithm of the ratio of these radii to determine the observational proxy $R_{\text{SF}}^{\texttt{FAST++}}$. We again describe alternative adaptations in Appendix~\ref{app:alternatives}. For both the inner and outer truncation radii, we adopt no adaptations as the specific star formation radial profiles are appropriate for use with the same formulation as used in the simulation, and as described in \citetalias{lawlor2026}.

With the fitted values and uncertainties, we draw 100 realizations for each radial profile (star formation, stellar mass, specific star formation) on a per-galaxy basis, where we assume the fitted values and corresponding uncertainties describe Gaussians. From the realizations of each profile, we compute the 
relevant observational morphological metrics for every realization, forming a distribution for each morphological metric per galaxy. We take the median across these values as the final value for the morphological metric, in order to incorporate information related to the uncertainties on the fitted values (stellar mass and star formation rate radial profiles). This process creates uncertainties for each metric for each galaxy that are similar to bootstrap errors.

\section{Results}\label{sec:results}

\begin{figure*}[t]
    \centering
    \includegraphics[width=\textwidth]{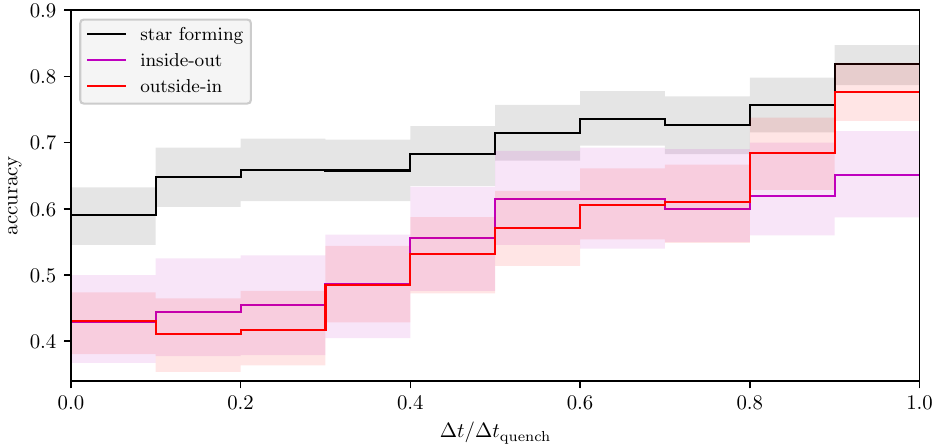}
    \caption{The accuracy of the machine learning-based nearest neighbor classification as a function of progress (time) through the quenching episode. For each population (star forming, black; inside-out quenching, magenta; outside-in quenching, red), we show median values across 1000 iterations as solid lines, with the corresponding shaded contours denoting the ${\pm} 1 \sigma$ range. For each iteration, a randomly selected sample of star forming galaxies is chosen to closely number the same number of quenching galaxies (inside-out + outside-in).}
    \label{fig:accuracy}
\end{figure*}

In Figure~\ref{fig:metric_comparison}, we show two dimensional histograms that compare the morphological metric observational proxies with the morphological metrics as used in the simulation. As above we take the difference of the observational proxies from the expected (simulation-based) values to describe the recovery using the median difference and scatter.

We find the best recovery when considering the concentration of star formation, $C_{\text{SF}}^{\texttt{FAST++}}$, with low scatter at all values. The median difference and scatter are $0.01$ and $0.05$, respectively, when considering all values, and ${<}0.01$ and $0.05$ when considering $C_{\text{SF}} \geqslant 0.5$. We likewise find excellent recovery when considering the size of the star forming disk relative to the stellar mass disk, $R_{\text{SF}}^{\texttt{FAST++}}$, with similarly low scatter at all values. The median difference and scatter are $0.01~\text{dex}$ and $0.14~\text{dex}$, respectively, for all values of $R_{\text{SF}}$, and ${<}0.01~\text{dex}$ and $0.13~\text{dex}$ for $R_{\text{SF}} \geqslant 0$. For the characteristic radii that trace sharp truncations of star formation, we find, in general, good recovery for both the inner truncation radius $R_{\text{inner}}^{\texttt{FAST++}}$, and the outer truncation radius $R_{\text{outer}}^{\texttt{FAST++}}$. For the inner truncation radius, away from the line of equality we can find small offsets, particularly at $R_{\text{inner}}^{\texttt{FAST++}}/R_{\text{e}} \approx 0$, that is commonly found for galaxies with $R_{\text{inner}}^{\text{TNG}}/R_{\text{e}} \lesssim 1$. We do not consider this small offset to be significant, as variations in the recovery of the radial profiles can directly impact the recoverability of the truncation radii. As well, we find that $|R_{\text{inner}}^{\text{TNG}} - R_{\text{inner}}^{\texttt{FAST++}}| \leqslant 0.2~R_{\text{e}}$ for 94\% of our sample galaxies. The median difference and scatter are $0.01$ and $0.13$ for all values of $R_{\text{inner}}$, and $0.06$ and $0.15$ for $R_{\text{inner}}/R_{\text{e}} > 0$. For the outer truncation radius, we find increasing scatter with increasing radii. This is not unexpected, as the outer regions of every galaxy will, in general, have lower signal-to-noise ratio data, and the recovered fitted radial profiles can be less accurate compared with the inner regions. Nonetheless, we find that our recovered outer truncation radii are accurate to within ${\pm} 1~R_{\text{e}}$ of $R_{\text{outer}}^{\text{TNG}}$ for 84\% of the sample, where this is largely driven by the scatter at large $R_{\text{outer}}$. The median difference and scatter are $-0.05$ and $0.58$, respectively, across all values of $R_{\text{outer}}$, but $-0.14$ and $0.49$ for $R_{\text{outer}} < 5$.

\begin{figure*}[t]
    \centering
    \includegraphics[width=\textwidth]{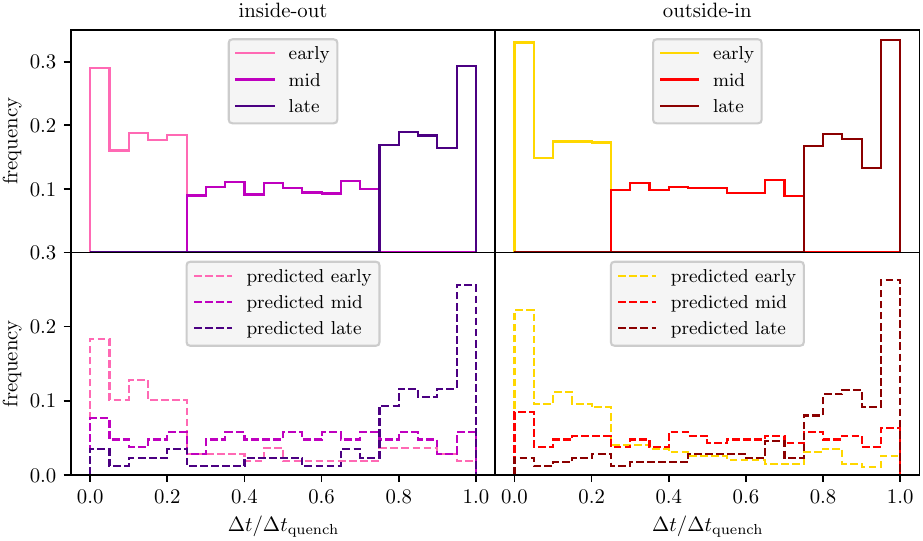}
    \caption{True (top) and predicted (bottom) quenching episode progress values for inside-out quenching galaxies (left) and outside-in quenching galaxies (right) using a machine learning-based \textit{k}-nearest neighbors classifier after 1000 iterations. For each panel, the histograms have been normalized so that the \textit{y}-axis can be read as frequency of occurrence. In each panel, populations are color coded by quenching episode progress epoch, where lighter colors correspond with earlier epochs and darker colors with later epochs. In the top panels, the boundaries that separate early-episode from mid-episode, and mid-episode from late-episode, are clearly seen at $\Delta t/\Delta t_{\text{quench}} = 0.25$ and $0.75$, respectively.}
    \label{fig:progress_recovery}
\end{figure*}

We next reproduce the analysis as completed in \citetalias{lawlor2026} that considers the feasibility of recovery the quenching population using machine learning classification. Specifically, we use a \textit{k}-nearest neighbors algorithm \citep[\textit{k}NN; e.g., ][]{fix1951,fix1989,cover1967} that is available in the package \texttt{scikit-learn} \citep{pedregosa2011}, and that uses the \textit{k}-nearest neighbors to a point in some hyperdimensional parameter space as samples to use for training. From these samples, a prediction can be made on the class or label for the given point of interest \citep{pedregosa2011}. We use our fitted observational proxies for the four morphological metrics and the true population labels (from \citetalias{lawlor2026}; i.e. star forming, inside-out, and outside-in) as the training set samples, and use the same 70\%$/$30\% training$/$testing split as in \citetalias{lawlor2026}. Following \citetalias{lawlor2026}, we use the three nearest neighbors, weighting the contribution of each neighbor based on their inverse distance, thereby ensuring that closer neighbors will have more influence compared to neighbors that are more distant. We additionally standardize$/$feature-scale the morphological metric proxies before classification.

In Figure~\ref{fig:accuracy}, we show the results of this analysis, where the progress through the quenching episode is binned into deciles. For each bin, we determine the accuracy of the predicted classification by comparison with the true population, adopted from the simulation and from \citetalias{lawlor2026}. The star forming population (black) is shown alongside the two quenching populations: inside-out (magenta) and outside-in (red). As each star forming galaxy has no quenching episode itself, we adopt the timestamps through the quenching episode from the corresponding quenching galaxy, and refer to this as the ``quenching episode'' for the star forming galaxy. For each progress bin$/$decile, we compute 1000 iterations where for each iteration we randomly select a number of star forming galaxies that closely numbers the same number of quenching (inside-out + outside-in) galaxies, and determine the correctly predicted classifications, forming an accuracy value. From these iterations, we then compute a median accuracy (solid lines) along with the corresponding ${\pm} 1 \sigma$ range for each population (shaded regions). In Figure~\ref{fig:accuracy}, we find that the accuracy for the star forming population increases throughout the quenching episode, starting from reasonable levels (${\sim}$60\%) at early times. Late in the quenching episode, the accuracy for the star forming population increases, exceeding 80\%. For the inside-out and outside-in populations, at early times the accuracy is low (${<} 0.5$), but by moving through the quenching episode, we see a general rise in the accuracy values, where at late times accuracies can exceed ${\sim}$75\% for the outside-in population. These results are similar to those found in \citetalias{lawlor2026}, though slightly lower at all times for all populations. This is not unexpected, given the degradation of the ``truth'' involved in producing the mock observations. However, even with slightly lower accuracy values at all epochs for the different populations, these results are encouraging and suggest that this methodology can properly classify the different populations, especially at later stages of quenching when the morphological differences are most pronounced.

Finally, we likewise reproduce the analysis as in \citetalias{lawlor2026} that examines the ability for a machine learning classification algorithm to correctly determine the stage at which a quenching galaxy is, when considering a given quenching population alone. In this scheme, we use the same \textit{k}-nearest neighbors classifier as employed above, and similarly use the three nearest neighbors, once again weighted by inverse distance. We first consider the inside-out population, and use the observational proxies for the morphological metrics and the corresponding values for the progress through the quenching episode as input to the classifier, where we bin the progress through the quenching episode into three epochs: early ($\Delta t/\Delta t_{\text{quench}} \leqslant 0.25$), mid-episode ($0.25 < \Delta t/\Delta t_{\text{quench}} < 0.75$), and late ($\Delta t/\Delta t_{\text{quench}} \geqslant 0.75$). We then predict the progress epoch, and repeat this analysis across 1000 iterations to produce distributions, from which we take the median. We then repeat this analysis considering only the outside-in population.

In Figure~\ref{fig:progress_recovery}, we show the results of this analysis as histograms where the progress through the quenching episode is binned and the \textit{y}-axis can be read as frequency. On the left, we show the results for the inside-out population, while on the right, we show the situation for the outside-in quenching galaxies. In the top row, we show the true distributions, while in the bottom row, we show the resulting predicted distributions. Considering first the top panel for the inside-out population, by construction the different epochs are perfectly separated. The true early galaxies have a peak at $\Delta t/\Delta t_{\text{quench}} = 0$, and the true late galaxies have a corresponding peak at $\Delta t/\Delta t_{\text{quench}} = 1$, where both features are due to the nature of the quenching episode progress definition, and all unique quenching galaxies will have both points by construction. We refer to \citetalias{lawlor2026}, from where we adopt the definition for this episode. In the bottom row we see that the predicted early episode galaxies are peaked at early values with a long tail to later values. Likewise, the predicted late episode galaxies are strongly peaked at late values with a tail to earlier values. The mid-episode galaxies reside in an intermediary space where they show no clear peak, though the mid-episode distribution shows tails into the early- and late-episode epochs as well. From this panel, it is clear that inside-out quenching galaxies that are truly early in their quenching episode are most commonly classified as early, while true late-episode galaxies are most often classified as late. Inside-out galaxies that are in the middle of their quenching episode are most difficult to distinguish, as they can appear similar to mature early-episode galaxies or to immature late-episode galaxies. When next considering the outside-in quenching population, we find an extremely analogous situation as for the inside-out population, where early- and late-episode galaxies are clearly morphologically distinct, but are closer to mid-episode galaxies, as expected. These results strongly resemble those found in \citetalias{lawlor2026}, and once again suggest that this methodology can be successfully applied to mock observations of simulated galaxies to predict the progress through the quenching episode for both inside-out and outside-in quenching galaxies.

\newpage
\section{Discussion}\label{sec:discussion}

We now discuss and interpret our results in the context of other studies presented in the literature, address our research questions as outlined above, and acknowledge limitations of the present work and future directions for follow-up.

By considering our results from the previous section, we are able to arrive at a number of fundamental conclusions based on the current study. The results have shown that by using mock CASTOR and NGRST imaging we are able to recover spatially resolved star formation rates and stellar masses, from which we construct radial profiles. These radial profiles can then be used to distinguish different modes of quenching, using the simple observational metrics introduced in \citetalias{lawlor2026}. When using machine learning classification algorithms on the observational proxies for the morphological metrics, we are able to accurately predict the true population (coming from the simulation) when considering normal star forming galaxies, inside-out quenching galaxies, and outside-in quenching galaxies, especially at late times. At early times, the quenching galaxies closely resemble star forming galaxies, and the morphological distinction is less clear. Finally, by considering a given quenching population alone, we can accurately distinguish galaxies that are in the early stages of their quenching episode from those that are in the late stages of quenching. Galaxies that are at intermediate stages are more difficult to accurately classify and can be misclassified as either early or late.

As presented in \citetalias{lawlor2026}, these morphological metrics evolve through the quenching episode, which we have recovered here. This evolution has strong distinguishing power, and can be used to well separate these populations. These findings suggest that this methodology is capable of capturing the true underlying population for star forming galaxies and galaxies that are in the stages of quenching, especially at later stages. This result has broad implications for the field of galaxy evolution, where studies can adopt these metrics and apply similar analyses to not only identify quenching galaxies based on morphology alone, but to identify galaxies that likely belong to a given quenching population.

\subsection{Comparison with Literature Results}

In the broader context of studies presented in the literature, we are not aware of any such study that completes a directly similar analysis as presented here. However, there has been much work done on investigating the morphology of quenching galaxies, particularly on how transition galaxies have morphologies intermediate between star forming galaxies and quiescent galaxies \citep[e.g.,][]{mendez2011,fang2013,almaini2017,rodriguez2019,matharu2020}. In addition, many works in the literature use morphological metrics to identify merging or interacting systems \citep[e.g.,][]{conselice2000,conselice2003,lotz2004,lotz2008,pawlik2016,powell2017,sazonova2024}, or even ram pressure-stripped systems, that are a clear sign of environmental factors that can quench galaxies in such dense environments \citep[e.g.,][]{holwerda2011,roberts2020,roberts2021,roman2021,krabbe2024}. Further, studies investigating the morphology of the emitting gas or dust emission have been used to probe the effect of the environment \citep[e.g.,][]{koopmann2004a} and distribution of star formation \citep[e.g.,][]{hall2018,diaz2020,shen2023} on quenching.

The studies most similar to the work presented here are investigations into the relationship and effect of quenching on galaxy morphology or the correlation between morphology and star formation activity as a function of redshift, to understand the morphological evolution of quenching galaxies \citep[e.g.,][]{wuyts2011,bell2012,lang2014,bluck2022}. Even still, these investigations consider large populations of quenching galaxies, and often explore the ensemble changes in morphology, given that they are unable to track the evolution of single galaxies given their observational nature.

\subsection{Surface Brightness Effects}

Before considering any applications, we discuss the effects of signal-to-noise ratio (SNR) as well as redshift dependence on the results. As mentioned above (Section~\ref{subsec:sed}), we generally find that with $\text{SNR} \gtrsim 3$ for a given annulus, that the stellar mass, star formation rate, metallicity, and dust are close in value to their counterparts as measurable or derivable from the simulation. In particular, we take the difference of the fitted values from the expected TNG values, and use the median difference and scatter for stellar mass (described above, see Section~\ref{subsec:sed}; $0.19~\text{dex}$ and $0.23~\text{dex}$) and star formation rate ($-0.13~\text{dex}$ and $0.46~\text{dex}$, respectively) as baseline values. Limiting our annuli to those with an \textit{H}-band $\text{SNR} \geqslant 10$, we find that the median difference and scatter for these values decrease: $0.17~\text{dex}$ and $0.20~\text{dex}$ for stellar mass, and $-0.12~\text{dex}$ and $0.38~\text{dex}$ for star formation rate, respectively. When further increasing the limiting \textit{H}-band SNR beyond ten, we find that the median stellar mass difference continues to decrease, as does the scatter. However, arbitrarily increasing the \textit{H}-band SNR does not oblige the \textit{UV}-band SNR to likewise increase, and we find no further improvements regarding the median star formation rate difference or scatter with increasing the \textit{H}-band SNR. However, if we instead consider limiting to \textit{UV}-band $\text{SNR} \geqslant 10$, then we find a median SFR difference of the same $-0.12~\text{dex}$, with a smaller scatter of $0.23~\text{dex}$. Further increasing the \textit{UV}-band SNR produces negligible improvements for the median difference, with only small improvements to the scatter.

Beyond individual annular values, the signal-to-noise ratio of the imaging has a direct impact on the quality of the recovered radial profiles. We again consider the \textit{H}-band SNR, but now investigate the number of annuli with $\text{SNR} \geqslant 10$ per galaxy, recalling that each profile has 20 annular bins. We split the entire sample by stellar mass, considering a low-mass population ($M_{*} < 10^{9.95}~M_{\odot}$) and a high-mass population ($M_{*} \geqslant 10^{9.95}~M_{\odot}$). The dividing stellar mass is arbitrary but was chosen to perfectly divide the sample in two. The average and median number of annuli with \textit{H}-band $\text{SNR} \geqslant 10$ for the low-mass sample are 14 and 13, respectively, while the corresponding number of annuli for the high-mass sample are 17.5 and 20, respectively. As expected, the higher mass galaxies have more high SNR annuli than their lower mass counterparts. If we instead consider a threshold SNR cut at $\text{SNR}_{\textit{H}\text{-band}} \geqslant 3$ to explore this lower SNR regime, we find that the average (median) number of annuli meeting the SNR threshold are 17 (18) and 18.8 (20) for the low- and high-mass populations, respectively. Taken together, these findings suggest that galaxies that do not have \textit{H}-band $\text{SNR} \geqslant 10$ for all of their annular bins are likely to have $\text{SNR} \geqslant 3$ for almost all of their annuli. This is especially true for the high-mass population, and remains true even for the low-mass population as well. This is reassuring given the desire to apply this methodology to different, particularly higher, redshifts.

For the current work, we produced mock CASTOR and NGRST images at a single characteristic intermediate redshift of $z = 0.5$, that is accessible to current and upcoming$/$proposed large galaxy surveys (e.g., Euclid, NGRST, LSST, and CASTOR). We additionally produced our mock images assuming the CASTOR Ultradeep survey and the NGRST HLWAS Deep tier and Ultradeep component. Given the effects of surface brightness dimming \citep[e.g.,][]{tolman1930,tolman1934,hubble1935}, moving to higher redshift will reduce the signal-to-noise ratio by a factor of $(1 + z)^{3}$, given the flux density per unit frequency units (i.e. janskys) used here \citep{whitney2020}. When applied to our existing results at $z = 0.5$ and investigating $z = 1$ as an example case, we find that the average and median number of annuli for the low-mass sample (as described above) with $\text{SNR}_{\textit{H}\text{-band}} \geqslant 3$ are 15 and 15, respectively. The high-mass sample has an average (median) number of annuli with $\text{SNR}_{\textit{H}\text{-band}} \geqslant 3$ of 18 (20). Compared with the situation at $z = 0.5$, this is a modest decrease and mostly strongly affects the low-mass sample, where the number of annuli with $\text{SNR}_{\textit{H}\text{-band}} \geqslant 3$ decreases by two on average$/$median. Using $\text{SNR}_{\textit{H}\text{-band}} \geqslant 10$ we find average (median) values of 11.3 (11) for the low-mass sample, and 15.8 (16) for the high-mass sample. This represents an inward shift to $4~R_{\text{e}}$ for the high-mass sample, while being limited to nearly ${\sim} 3~R_{\text{e}}$ for the low-mass galaxies. When additionally considering the redshift dependence of angular size, we find that by moving to $z = 1$ from $z = 0.5$, we would obtain mock observations that are 76.4\% the size of our fiducial images. Given the size-mass relation \citep[e.g.,][]{shen2003,vanderwel2014,lange2015,mowla2019}, this effect would be most detrimental to the lowest mass galaxies, making their mock observations even smaller than presently. Though we expect that the blurring effect of the point spread function is small (see below), images that are comparable in size to the point spread function would become problematic for accurately recovering their stellar mass and star formation rate radial profiles. However, when considering galaxies in our sample with $10^{9.5} \leqslant M_{*}/M_{\odot} < 10^{9.75}$, we find that the median mock image size is 101 pixels square at $z = 0.5$, resulting in $z = 1$ images of median size $\sim$77 pixels square, which is sufficient to complete the analysis as described in this work.

These findings suggest that the methodology presented in the current work should be suitable for large galaxy surveys, but that care must be taken when adapting the technique, as lower mass galaxies at higher redshift may not have sufficient signal-to-noise extending to ${\sim} 5~R_{\text{e}}$. However, these findings also indicate that even intermediate mass galaxies ($M_{*} \approx 10^{10}~M_{\odot}$) should have profiles with sufficient signal-to-noise to carry out the kind of analysis described above.

\subsection{Application to Future CASTOR and NGRST Surveys}

We now discuss the feasibility of applying our methodology to samples of galaxies in the real Universe, by first considering the expected number of galaxies in the proposed CASTOR surveys and the upcoming NGRST surveys.

In \citetalias{lawlor2026}, we calculated the galaxy stellar mass function for our simulated sample of quenched galaxies, given the total simulation volume of TNG50 ($35/h~\text{Mpc}$ on a side, where $h = 0.6774$). We additionally compared our calculated stellar mass function with stellar mass functions for local ($0.004 < z < 0.08$) red galaxies \citep{baldry2004} in the Sloan Digital Sky Survey \citep[SDSS;][]{york2000}, and local ($z < 0.06$) red galaxies \citep{baldry2012} in the Galaxy And Mass Assembly survey \citep[GAMA;][]{driver2011}. We note that \citet{baldry2004,baldry2012} define red galaxies via location in $u-r$ color--magnitude space. In \citetalias{lawlor2026}, we found good agreement between the shapes of our measured galaxy stellar mass function and those of \citet{baldry2004,baldry2012}, but with a noticeable shift to lower values for all stellar masses for our empirical function, reflecting the more restrictive definition of quenched galaxy. Following this, by fitting a Schechter function \citep{schechter1976} to our results from \citetalias{lawlor2026}, we can make predictions on the abundances of quenched galaxies that satisfy our (conservative) definition of a quenched galaxy (see Section~\ref{subsec:tng}).

\begin{figure}
    \centering
    \includegraphics[width=\columnwidth]{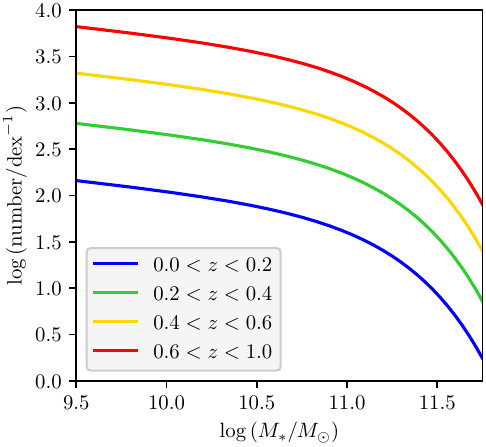}
    \caption{Expected number of galaxies per stellar mass bin for different redshift shells, determined using the fitted Schechter function as shown in Equation~(\ref{eq:smf}) and the total volume of space within each redshift shell. Each redshift shell is shown with distinct colors.}
    \label{fig:expected}
\end{figure}

In the present work we used a single Schechter function of the form \citep{weigel2016}:
\begin{align}\label{eq:smf}
    n_{\text{galaxies}} &= \Phi(M) \mathrm{d}\log{M} \nonumber \\
    &= 
    \ln{(10)}~\Phi^{*}~e^{-10^{\log{M} - \log{\mathcal{M}^{*}}}} \nonumber \\
    & \hspace{4mm} \times \left( 10^{\log{M} - \log{\mathcal{M}^{*}}} \right)^{\alpha + 1} \mathrm{d}\log{M},
\end{align}
where $n_{\text{galaxies}}$ is the number density of galaxies in a mass bin of $\mathrm{d}\log{M}$, $M$ is the stellar mass, $\mathcal{M}^{*}$ is the characteristic mass that describes the ``knee'' of the Schechter function, $\alpha$ is the power law slope for masses smaller than $\mathcal{M}^{*}$, and $\Phi^{*}$ describes the overall normalization and is the number density at $\mathcal{M}^{*}$. We fit our calculated stellar mass function for galaxies with $9.5 \leqslant \log{(M_{*}/M_{\odot})} \leqslant 11.75$ with the above, and find best fit values of $\log{(\mathcal{M}^{*}/M_{\odot})} = 11.22$, $\alpha = -1.21$, and $\Phi^{*}/10^{-3} = 0.47$. We next consider the area of the CASTOR Ultradeep survey ($1~\text{deg}^{2}$, that should overlap with the deep NGRST surveys; see above) and representative low- and intermediate-redshifts that will be accessible to CASTOR$/$NGRST and determine the total volume of space within these redshift shells. We then calculate expected galaxy counts based on our fitted stellar mass function for our quenched population, and show our findings in Figure~\ref{fig:expected}.

\begin{figure*}
    \centering
    \includegraphics[width=\textwidth]{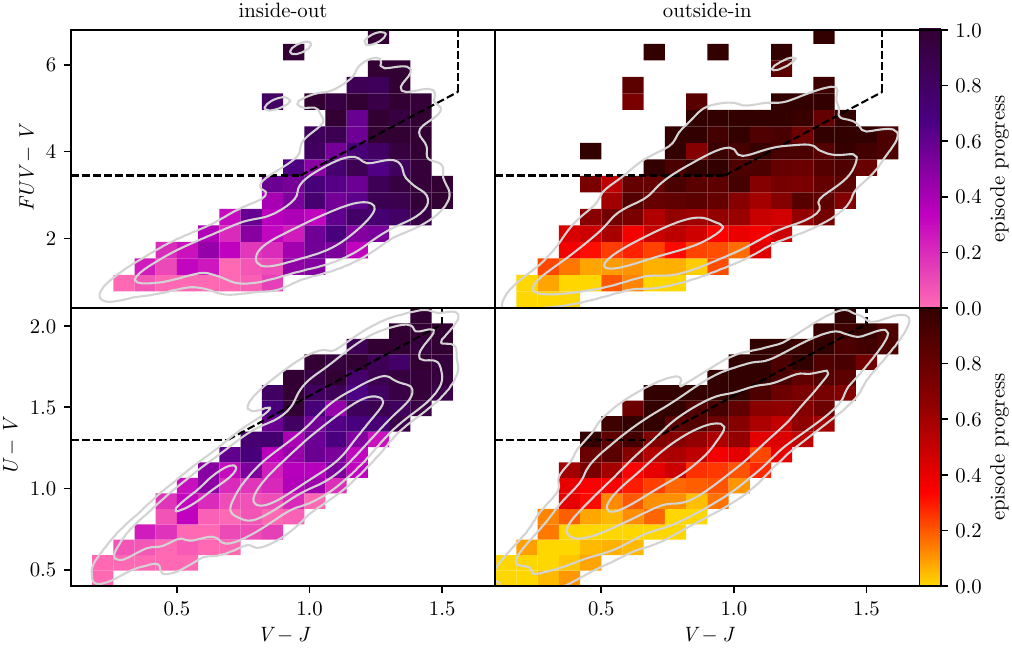}
    \caption{Color--color diagrams showing the median progress through the quenching episode per bin for inside-out (left) and outside-in (right) quenching galaxies, color-coded by progress through the quenching episode. Earlier times in the quenching episode are colored more lightly (pink and gold, respectively), while those later in the quenching episode are darker (dark magenta and dark red, respectively). The upper panels show the location of galaxies in \textit{FUV}--\textit{V}, \textit{V}--\textit{J} color space, while the lower panels show the standard \textit{UVJ} diagram. In the upper panels, we additionally show the division of \citet{leja2019b} for star forming and quiescent galaxies (black dashed line) while the lower panels show the division of \citet{muzzin2013}. In all panels, contours that denote the location and density of the respective quenching population are shown in silver.}
    \label{fig:fuvvj_and_uvj}
\end{figure*}

Considering Figure~\ref{fig:expected}, we expect that more than 99,000 (${\pm}$4800) galaxies with $M_{*} > 10^{9.5}~M_{\odot}$ should be within the survey footprint at $z < 1$ assuming a bin width of $0.1~\text{dex}$, while there should be more than 27,000 (${\pm}$2500) galaxies with $M_{*} > 10^{10.5}~M_{\odot}$. This amounts to a large sample of quenching galaxies to investigate and is bolstered by the number of galaxies at higher redshifts, given the large corresponding volume. Indeed, when considering the highest redshift bin in Figure~\ref{fig:expected}, we expect to find more than 69,000 (${\pm}$3000) of the $M_{*} > 10^{9.5}~M_{\odot}$ galaxies between $z = 0.6$ and $z = 1$, while there should be some 19,000 (${\pm}$1500) galaxies with $M_{*} > 10^{10.5}~M_{\odot}$. This subpopulation may prove critical in applying this type of morphological investigation to higher redshift galaxies, given the discussion of signal-to-noise and surface brightness dimming in the proceeding section. At intermediate redshifts like the one considered in this work ($z \approx 0.5$), we anticipate some 22,000 (${\pm}$1000) galaxies with $M_{*} > 10^{9.5}~M_{\odot}$, and more than 6000 (${\pm}$500) quenching galaxies with $M_{*} > 10^{10.5}~M_{\odot}$. Below $z \approx 0.4$, we expect some 7800 (${\pm}$400) galaxies above $M_{*} = 10^{9.5}~M_{\odot}$, and even 2100 (${\pm}$200) galaxies at higher stellar masses ($M_{*} > 10^{10.5}~M_{\odot}$). These findings indicate that with surveys such as the CASTOR Ultradeep survey alongside the Ultradeep component of the NGRST HLWAS, many thousands of these types of quenching galaxies should be visible, and should be ideal candidates for applying the type of methodology and morphological study as presented in this work.

Beyond the expected number of galaxies, there should also be several massive ($M > 10^{14}~M_{\odot}$) galaxy clusters at the redshifts under consideration. Using similar redshift slices as above and the area of the NGRST HLWAS Deep tier ($19.2~\text{deg}^2$), we expect ${\sim} 10$, ${\sim} 10$, and ${\sim} 3$ clusters for $0 \leqslant z \leqslant 0.2$, $0.2 \leqslant z \leqslant 0.4$, and $0.4 \leqslant z \leqslant 0.6$, respectively \citep{mak2011}. The cluster galaxies will prove invaluable for understanding the effect of the environment.

In addition to the many thousands of galaxies and handful of massive galaxy clusters that should be within such fields, we also consider the locations of such galaxies in color--color space, a popular technique for separating star forming galaxies from quiescent galaxies \citep[e.g.,][]{strateva2001,daddi2004,williams2009,ilbert2013}. We use the standard \textit{UVJ} diagram \citep[e.g.,][]{williams2009} as well as the \textit{FUV}--\textit{V}--\textit{J} diagram \citep[e.g.,][]{leja2019b}.

In Figure~\ref{fig:fuvvj_and_uvj}, we show our findings for the inside-out (left) and outside-in (right) quenching populations, where the top panels show the \textit{FUV}--\textit{V}--\textit{J} diagrams and the bottom panels show the \textit{UVJ} diagrams. We bin galaxies by location in color--color space, and bins are color-coded according to the median progress through the quenching episode for galaxies within that bin. Using the mean progress as compared to the median progress produces negligible differences. Lighter colors correspond to earlier times, and darker colors to later. In the \textit{FUV}--\textit{V}--\textit{J} diagrams we show the division (black dashed line) of \citet{leja2019b} that best separates star forming galaxies from quiescent galaxies based on a cut in sSFR for galaxies in 3D-HST \citep{brammer2012}, while in the \textit{UVJ} diagrams we show similar divisions from \citet{muzzin2013} for $z < 1$ galaxies from COSMOS$/$UltraVISTA \citep{mccracken2012}. In all panels, we show unfilled contours that describe the density and location of the respective quenching population with silver. Though we do not show contours that describe the density and location of normal star forming galaxies for visual clarity, these populations peak in density close to the densest region for the quenching populations, but generally with an offset to lower values of \textit{V}--\textit{J} and \textit{FUV}--\textit{V} (upper panels) or \textit{U}--\textit{V} (lower panels), and additionally do not occupy the quiescent region in any panel. Considering Figure~\ref{fig:fuvvj_and_uvj}, as we expect, we see that quenching galaxies that are in the early stages of quenching (light colors) generally reside within the locus of the star forming region, but as galaxies begin to, and continue to quench, they move to the quiescent region for each respective diagram (dark colors). These findings indicate that by carefully using such color--color diagrams, one can select quenching galaxies that are further along in their quenching process on the basis of synthesized colors (especially in the passive regions), though we discuss limitations below.

Taken together, these findings demonstrate the feasibility of first finding, then correctly identifying such quenching galaxies within fields such as the CASTOR Ultradeep field and the deep fields of NGRST's HLWAS. We expect that when applied to such datasets, the methodology outlined in this paper will prove useful for identifying and distinguishing different populations of quenching galaxies to great success, using morphological considerations alone. With such samples available, additional follow-up can focus on various aspects of each population, such as environment (both local and large-scale), interactions and mergers, prevalence and strength of active galactic nucleus feedback, and kinematics, for example.

\subsection{Limitations and Future Directions}

When considering the limitations of the present work, we find a few points worth commenting on, and discuss each as it arises chronologically in the methodology presented above.

The first concerns how representative simulated galaxies are of real galaxies. Previous studies have proposed that the active galactic nuclei feedback model is strong in TNG when compared to observations \citep[e.g.,][]{terrazas2020,donnari2021,voit2024,wright2024}, resulting in inside-out quenching that is pronounced compared to observations. Additional work has suggested that environmental gas removal may be overly strong \citep[e.g.,][]{diemer2019,stevens2019,stevens2021,stevens2023}, resulting in dramatic outside-in quenching. Therefore applying the morphological metrics to real galaxies must be done carefully, especially when considering $R_{\text{inner}}$ and $R_{\text{outer}}$. However, these metrics are less important compared to $C_{\text{SF}}$ and $R_{\text{SF}}$ when identifying quenching galaxies (see Appendix~D of \citetalias{lawlor2026}).

The next concern is the similarity between the prescription for generating the mock observations, the \textsc{galaxev} pipeline described above, and the setup for the SED fitting process, as both the \textsc{galaxev} pipeline and the SED fitting code \texttt{FAST++} rely on the models of \citet{bruzual2003}. In reality, galaxies have complex star formation histories and metallicity and dust distributions, which means that our ability to recover star formation rates here is a best-case scenario. Even so, one could instead generate the mock observations using different libraries such as Starburst99 \citep{leitherer1999}, those of \citet{maraston2005}, FSPS \citep{conroy2009,conroy2010}, or BPASS \citep{eldridge2017}, or using different pipelines \citep[e.g.,][]{fortuni2023,janulewicz2025,lovell2025,zhou2025,roper2026}, though the use of these alternative libraries or pipelines is outside the scope of this work.

Beyond these initial considerations, the next principle consideration is an improvement regarding a more physically realistic dust setup. A natural approach is to use a full radiative transfer code such as SKIRT \citep{baes2003,baes2011,baes2015,camps2015,camps2020} or \textsc{powderday} \citep{narayanan2021}. Indeed, many papers in the literature have used SKIRT with TNG galaxies for various investigations \citep[e.g.,][]{huertas2019,whitney2021,trcka2022,costantin2023,baes2024,bottrell2024}. However, following \citet{rodriguez2019} and their considerations, in this work we took a simpler approach by using a foreground dust screen based on \citet{calzetti2000}. As mentioned above, we did initially pursue using SKIRT but found for our purposes that the results were not sufficiently different to warrant the extra computation. In either event, the effect of dust will most strongly affect young stellar populations, and when considering galaxies in the later stages of quenching this effect should be small. An additional minor consideration is the blurring effect of the point spread function when producing our mock observations. Given the PSF used above (FWHM of $0 \farcs 15$), we expect the effect of the PSF should be small but non-negligible (see, for example, the last row of Figure~\ref{fig:mock_images}). Indeed, when we compared our SED-fitted values for stellar mass and star formation rate that are based on the PSF-convolved mock observations with their simulation-based counterparts (Figure~\ref{fig:comparison}), we found small differences that are not systematic, further supporting our conclusion that the effect of the PSF is small. However, this is based on the PSF used in the current work, and similar conclusions may not hold for observations with a much larger point spread function.

Alternative approaches or implementations in the current work could include using a sophisticated SED fitter with Bayesian inference \citep[like \textsc{Bagpipes};][]{carnall2018,carnall2019}, or even a fitter that can use or uses fully non-parametric star formation histories, like \textsc{prospector} \citep{leja2017,leja2019a,johnson2021} or Dense Basis \citep{iyer2017,iyer2019}, respectively. In the present work, we have opted to use the relatively simpler and more straight forward grid fitter \texttt{FAST++}, and we do not anticipate that using an alternative SED fitter should have a large effect on our presented results. In fact, though we have used a relatively simpler fitter, we have still recovered key parameters such as stellar mass, star formation rate, metallicity, and dust, when comparing the fitted results with similar properties within the simulation (see Figure~\ref{fig:comparison} and surrounding discussion).

An additional approach is to do full pixel-by-pixel SED fitting and subsequently bin the fitted results together to recover the annular bins. We briefly explored this option but issues with low surface brightness and low signal-to-noise ratio data (especially in the outskirts of many galaxies) prevented this approach from accurately recovering the expected star formation history parameters like stellar mass and star formation rate. Beyond pixel-by-pixel fitting, one could also first perform voronoi tesselation using such software \citep[e.g., VorBin;][]{cappellari2003}, and subsequently complete SED fitting on the voronoi-binned photometry, or alternatively use a fitter that accounts for the similarity in the shapes of the SEDs when binning pixels together \citep[e.g., \texttt{piXedfit};][]{abdurrouf2021}. Both of these latter approaches hold promise, though the issue with the low signal-to-noise data in outer regions will still provide some difficulties. In either event, such approaches are outside the scope of this work.

When considering the ability to identify quenching galaxies in a representative sample from a large galaxy survey, we face challenges due to the dominant star forming population. The ability to find ``needles in a haystack'' using our approach needs further refinement, and may require additional information beyond the morphological metrics alone. For instance, including the integrated star formation rate or the distance from the star forming main sequence could prove vital for properly identifying quenching galaxies. Indeed, in the above analysis where we showed we can distinguish different populations of quenching galaxies from normal star forming galaxies, we used a balanced sample of star forming galaxies and quenching (inside-out + outside-in) galaxies. We note that in the star forming region of the \textit{UVJ} diagrams above (Figure~\ref{fig:fuvvj_and_uvj}), the star forming population will be heavily dominant, so we expect that our recovery would have many false positives, thus proving difficult to truly identify early quenching galaxies. However, Figure~\ref{fig:fuvvj_and_uvj} does show that among a \textit{UVJ}-passive sample there is minimal contamination from star forming galaxies, and we can therefore apply this technique to identify later inside-out and outside-in quenching galaxies.

Regarding future directions, the next step is to apply the methodology outlined here to populations of galaxies in the real Universe. To facilitate this, we will aim to use high spatial resolution data with deep integration times. In order to make additional comparisons across various cosmic environments, the ideal dataset would include observations of both field and cluster galaxies, as well as those in intermediate density environments between these two extremes. Promising datasets to consider include large campaigns completed with HST, given the excellent spatial resolution and wavelength coverage afforded by such observations. Programs such as the Cosmic Assembly Near-infrared Deep Extragalactic Legacy Survey \citep[CANDELS;][]{grogin2011,koekemoer2011}, the Cluster Lensing And Supernova Survey with Hubble \citep[CLASH;][]{postman2012}, the Hubble Frontier Fields \citep[HFF;][]{lotz2017}, the Reionization Lensing Cluster Survey \citep[RELICS;][]{coe2019}, or the COSMOS Legacy UV-Optical Treasury Campaign with Hubble (CLUTCH; HST program ID 17802; PI: J.~Kartaltepe) would satisfy some or all of the above desirable criteria. This is the scope of our next investigation (Lawlor-Forsyth et al. 2026, in preparation).

\section{Summary and Conclusion}\label{sec:summary}

In this paper, we have investigated the feasibility of recovering morphological metrics for mock observations of galaxies present in the high resolution simulation TNG50. We first adopt the sample of 361 quenched galaxies from TNG50 as initially determined in \citetalias{lawlor2026}, and consider all time points within each galaxies' respective quenching episode, alongside normal star forming galaxies that have a similar stellar mass as the quenching galaxies. We arrive at a sample of 5365 quenching galaxies and 64,128 unique star forming galaxies. For each galaxy, we produce mock images that are based on proposed or upcoming galaxy surveys with CASTOR \citep{cote2025} and NGRST \rotacp, and bin the photometry into concentric annuli. Our main findings are as follows:
\begin{enumerate}[noitemsep]
    \item By employing a hybridized parametric star formation history (e.g., Figure~\ref{fig:sfhs_seds}) that can capture both the broad scale evolution of long-lived stellar populations and provide flexibility for bursts or truncations of star formation over short timescales ($100~\text{Myr}$), we are able to accurately recover key parameters like stellar mass and star formation rate when compared to the corresponding simulation parameters (Figure~\ref{fig:comparison}).
    
    \item By reconstructing radial profiles of stellar mass and star formation rate through the use of the annular fits, we are able to produce radial profiles that are close matches to their simulation counterparts, where the accuracy of the recovered profile depends on surface brightness and signal-to-noise ratio (Figure~\ref{fig:profiles}).
    
    \item Using the recovered radial profiles and the adapted observational proxies for the morphological metrics, the concentration of star formation and the ratio of disk sizes are most accurately recovered, while the characteristic radii that trace sharp truncations in star formation are slightly more difficult to recover, especially $R_{\text{outer}}$ (Figure~\ref{fig:metric_comparison}).
    
    \item Machine learning classification algorithms can separate star forming galaxies from quenching galaxies, and further, can separate inside-out quenching galaxies from outside-in quenching galaxies, especially at later times in the quenching episode, when the morphology of these different populations of galaxies is most distinct (Figure~\ref{fig:accuracy}).
    
    \item Machine learning classifications can also be used well to predict the epoch of progress through the quenching episode for the two main populations of quenching galaxies, when each is considered independently. Early-episode galaxies are most easily distinguished from late-episode galaxies, while mid-episode galaxies can be more difficult to distinguish given their resemblance to mature early galaxies and immature late galaxies (Figure~\ref{fig:progress_recovery}).
\end{enumerate}

These findings as described above suggest that the observational adaptations of the morphological metrics can be successfully applied to mock observations of simulated galaxies, and can be used to recover information about the intrinsic state of quenching galaxies based only on morphology alone. Caution must be taken when applying such metrics to galaxies with properties (e.g., surface brightness, angular size, etc.) that differ from those in the simulated images considered here. Our next step in this series of investigations is to apply our findings to observations of galaxies in the real Universe, using high spatial resolution observations with deep integration times. We will consider galaxies in dense environments and those residing in the field to enable comparisons between different populations, thereby investigating the effect of the environment.

\section*{Acknowledgments}

C.L.F. thanks Corentin Schreiber for help with the fitting software \texttt{FAST++}, and Pat C{\^o}t{\'e} and Tyrone Woods for discussion regarding CASTOR. The authors would like to acknowledge research grant funding that has enabled this research, including the support of the Natural Sciences and Engineering Research Council of Canada (NSERC) grants RGPIN-2018-03820 and RGPIN-2024-03878 for C.L.F. and M.L.B. S.L.M. acknowledges support from Science and Technology Facilities Council grants ST$/$W000946$/$1 and ST$/$Y000692$/$1. G.H.R. acknowledges the support of National Science Foundation grants AST-2206473 and AST-2308126, and grant 80NSSC21K0641 issued through the NASA Astrophysics Data Analysis Program (ADAP).

The IllustrisTNG simulations were undertaken with compute time awarded by the Gauss Centre for Supercomputing (GCS) under GCS Large-Scale Projects GCS-ILLU and GCS-DWAR on the GCS share of the supercomputer Hazel Hen at the High Performance Computing Center Stuttgart (HLRS), as well as on the machines of the Max Planck Computing and Data Facility (MPCDF) in Garching, Germany.

The authors acknowledge the use of the Canadian Advanced Network for Astronomy Research (CANFAR) Science Platform operated by the Canadian Astronomy Data Center (CADC) and the Digital Research Alliance of Canada (DRAC), with support from the National Research Council of Canada (NRC), the Canadian Space Agency (CSA), CANARIE, and the Canadian Foundation for Innovation (CFI).

This research has made use of the Astrophysics Data System, funded by NASA under Cooperative Agreement 80NSSC25M7105, as well as \texttt{TOPCAT},\footnote{\href{https://www.star.bris.ac.uk/~mbt/topcat}{https:$//$www.star.bris.ac.uk$/{\sim}$mbt$/$topcat}} an interactive graphical viewer and editor for tabular data \citep{taylor2005}, in addition to \texttt{Astropy},\footnote{\href{https://www.astropy.org}{https:$//$www.astropy.org}} a community-developed core Python package and an ecosystem of tools and resources for astronomy \citep{astropy2013,astropy2018,astropy2022}, and \texttt{Photutils}, an Astropy package for detection and photometry of astronomical sources \citep{bradley2025}.

\vspace{0.3cm}
\textit{Software:} \texttt{Astropy} \citep{astropy2013,astropy2018,astropy2022}, \texttt{FAST++} \citep{kriek2009,schreiber2018a}, \texttt{h5py} \citep{collette2023}, \texttt{Matplotlib} \citep{hunter2007}, \texttt{NumPy} \citep{harris2020}, \texttt{Photutils} \citep{bradley2025}, \texttt{scikit-learn}\footnote{\href{https://scikit-learn.org}{https:$//$scikit-learn.org}} \citep{pedregosa2011}, \texttt{SciPy} \citep{virtanen2020}, \texttt{TOPCAT} \citep{taylor2005}

\appendix

\section{Alternative Morphological Metric Adaptations}\label{app:alternatives}

\begin{figure*}[t]
    \centering
    \includegraphics[width=\textwidth]{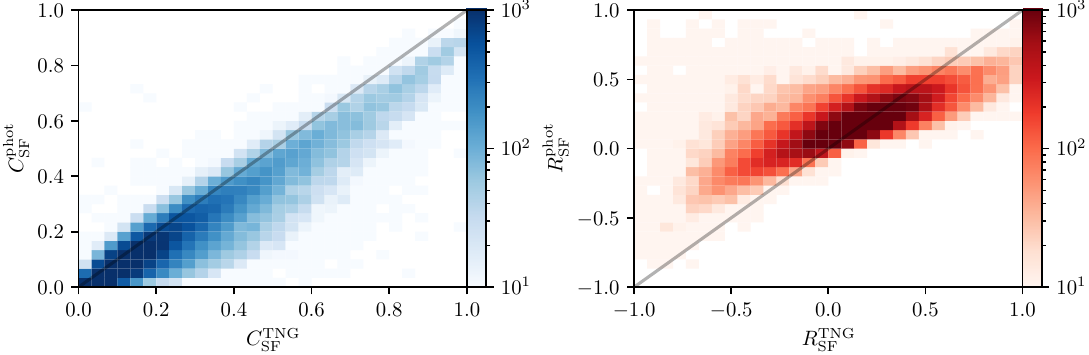}
    \caption{Similar to the top panels of Figure~\ref{fig:metric_comparison}, comparisons of the alternative observational proxies of the morphological metrics using only the photometry as a function of the morphological metrics as computed using TNG. In both panels, we show 2D histograms where the color scale describes the density of points, and a line of equality is shown in gray.}
    \label{fig:alternative}
\end{figure*}

Here, we present and discuss alternative adaptations to the morphological metrics, as initially described in Section~\ref{subsubsec:adaptations}. In the course of identifying appropriate adaptations that must be made for the morphological metrics from \citetalias{lawlor2026}, given that we are applying these metrics to mock observations, we investigated several versions of some of the morphological metrics, namely $C_{\text{SF}}$ and $R_{\text{SF}}$. In the body of the text we present our preferred observational adaptations that are based on interpolations of cumulative mass profiles for both metrics. These adaptations are preferred as they best recover the morphological metrics as applied in the simulation, as described in \citetalias{lawlor2026}. However, we also examined the recovery using versions based solely on the photometry, without using the fitted SED results.

For $C_{\text{SF}}$, we considered the CASTOR \textit{UV} image as a proxy for the star formation rate, and used the luminosity within $1~\text{kpc}$ (projected) and the total integrated luminosity, taking the ratio as $C_{\text{SF}}$. We show the results of this investigation in the left panel of Figure~\ref{fig:alternative}. We find a systematic offset to lower values for the photometry-based values, where at a fixed value of $C_{\text{SF}}^{\text{TNG}}$ the value of $C_{\text{SF}}^{\text{phot}}$ is almost exclusively lower. However, we do find that $C_{\text{SF}}^{\text{phot}}$ nearly covers the entire range of $C_{\text{SF}}^{\text{TNG}}$ values, as we might expect. The median difference and scatter are $0.03$ and $0.07$, respectively, when considering all values, and $0.03$ and $0.06$ when considering $C_{\text{SF}}^{\text{TNG}} \leqslant 0.5$. These results indicate that solely relying on the CASTOR \textit{UV} photometry to form an observational proxy for $C_{\text{SF}}$ is an appropriate starting point, but completing SED fitting is crucial to recover the associated star formation rate (and profile) from which $C_{\text{SF}}$ is best derived.

For $R_{\text{SF}}$, we similarly considered the CASTOR \textit{UV} image as a proxy for the star formation rate and the NGRST \textit{H}-band image as a proxy for the stellar mass. For both images, we determined the radius enclosing half of the total flux, taking these radii as proxies for the effective radii, respectively. Finally, we use the logarithm of the ratio of these radii as $R_{\text{SF}}$. As above, we show the results of this alternative parametrization in the right panel of Figure~\ref{fig:alternative}. We find that using the photometry alone produces $R_{\text{SF}}^{\text{phot}}$ values with less range compared to their $R_{\text{SF}}^{\text{TNG}}$ counterparts, where the values for $R_{\text{SF}}^{\text{phot}}$ almost exclusively lie between $-0.5$ and $0.5$. Taken alone, this reduced range is not critically detrimental, but in addition, there exists an apparent rotation of the shape of the distribution, where values of $R_{\text{SF}}^{\text{phot}}$ at low $R_{\text{SF}}^{\text{TNG}}$ are above the line of equality (gray), while those at high values of $R_{\text{SF}}^{\text{TNG}}$ are below the line of equality. The median difference and scatter are $-0.03~\text{dex}$ and $0.19~\text{dex}$, respectively, for all values, $-0.20~\text{dex}$ and $0.21~\text{dex}$ for $R_{\text{SF}}^{\text{TNG}} < 0$, and $0.01~\text{dex}$ and $0.12~\text{dex}$ for $R_{\text{SF}}^{\text{TNG}} \geqslant 0$. As above, these results indicate that completing SED fitting is necessary in order to best find observational proxies for $R_{\text{SF}}$ with a similar range and with small non-systematic offsets.

\section*{ORCID iDs}
\begingroup
\raggedright
Cameron~Lawlor-Forsyth \orcidlink{0000-0002-2958-0593} \\\href{https://orcid.org/0000-0002-2958-0593}{https:$//$orcid.org$/$0000-0002-2958-0593}\par
Michael~L.~Balogh \orcidlink{0000-0003-4849-9536} \href{https://orcid.org/0000-0003-4849-9536}{https:$//$orcid.org$/$0000-0003-4849-9536}\par
Sean~L.~McGee \orcidlink{0000-0003-3255-3139} \href{https://orcid.org/0000-0003-3255-3139}{https:$//$orcid.org$/$0000-0003-3255-3139}\par
Gregory~H.~Rudnick \orcidlink{0000-0001-5851-1856} \\\href{https://orcid.org/0000-0001-5851-1856}{https:$//$orcid.org$/$0000-0001-5851-1856}
\endgroup

\bibliographystyle{aasjournal}
\bibliography{references}

\end{document}